%% file: pmain.tex
\def\dtitle{Transverse momentum spectra and nuclear modification factors of charged particles in \xexe collisions \\ at $\mathbf{\sqrsn=~5.44}~\text{TeV}$}
\def\stitle{$R_{\rm AA}$ of charged particles in \xexe collisions at $\sqrt{s_{\rm NN}}=5.44$ TeV} 
\documentclass[ALICE,manyauthors,12pt]{cernphprep}
\usepackage[comma,square,numbers,sort&compress]{natbib}
\usepackage{hyperref}
\usepackage{float}
\usepackage{graphicx}  
\usepackage{dcolumn}   
\usepackage{bm}        
\usepackage{amssymb}   
\usepackage{amsfonts}
\usepackage{graphics}
\usepackage{grffile}   
\usepackage{epsfig}
\usepackage[loose]{units}
\usepackage[usenames,dvipsnames]{xcolor}
\usepackage[normalem]{ulem} 
\usepackage{color}
\usepackage[utf8]{inputenc}
\usepackage[T1]{fontenc}
\usepackage{subfigure}

\input{newcommands.tex}

\usepackage{lineno}
\begin{document}
\begin{titlepage}
\PHyear{2018}
\PHnumber{112} 
\PHdate{May 9}   
\title{\dtitle}
\ShortTitle{\stitle}
\Collaboration{ALICE Collaboration%
         \thanks{See Appendix~\ref{app:collab} for the list of collaboration members}}
\ShortAuthor{ALICE Collaboration} 
\begin{abstract}
Transverse momentum (\pt) spectra of charged particles at mid-pseudorapidity in \xexe collisions at $\sqrsn~=~5.44~\tev$ 
measured with the ALICE apparatus at the Large Hadron Collider are reported.
The kinematic range $0.15 < \pt < 50\ \gevc$ and $|\eta| < 0.8$ is covered. Results are presented in nine classes of collision centrality in the 0--80\% range. 
For comparison, a \pp reference at the collision energy of $\sqrs$~=~5.44~\tev is obtained by interpolating between existing \pp measurements at  $\sqrs$~=~5.02 and 7~\tev.
The nuclear modification factors in central \xexe collisions and \pbpb collisions at a similar center-of-mass energy of $\sqrsn$~=~5.02~\tev, and in addition at 2.76~\tev,
at analogous ranges of charged particle multiplicity density \avg{\dnchdeta} show a remarkable similarity at $\pt > 10$~\gevc.
The centrality dependence of the ratio of the average transverse momentum \mpt  in \xexe collisions over \pbpb collision at $\sqrs$~=~5.02~\tev is compared to hydrodynamical model calculations.
\end{abstract}
\end{titlepage}
\newpage
\setcounter{page}{2}
\section{Introduction}
\label{sec:intro}
Transverse momentum (\pt) spectra of charged particles carry essential information about the high-density deconfined state of strongly-interacting matter
commonly denoted as quark-gluon plasma, that is formed in
high-energy nucleus-nucleus~(A--A) collisions~\cite{Muller:2012zq}. Relativistic hydrodynamics is able to model the evolution of this medium~\cite{Heinz:2013th,Bernhard:2016tnd}.

At low to intermediate \pt, typically in the range of up to 10~\gevc, charged particle production is governed by the collective expansion of the system, which is observed in the 
shapes of single-particle transverse-momentum spectra~\cite{Adam:2016tre,Acharya:2018qsh} and multi-particle correlations~\cite{Heinz:2013th}. However, there is presently an intense debate as to whether the strikingly similar signatures observed in small collision systems (\pp and \pA) are also of hydrodynamical origin~\cite{Adam:2016dau,Ortiz:2013yxa,ABELEV:2013wsa,Khachatryan:2014jra,Chatrchyan:2013nka,Khachatryan:2015waa,Aad:2013fja,Dusling:2013oia,Blok:2017pui}.
A key ingredient of calculations in relativistic  hydrodynamics is the initial energy density~\cite{Heinz:2013th,Adare:2015bua, Adam:2016thv}. 
The number of produced particles and the volume of the medium are approximately proportional to the number of nucleons $\Npart$ that participate in the collision~\cite{Adam:2015vna, ALICEpubCent,xexe-cent}.
Thus, the particle density per unit volume is roughly independent of $\Npart$.  As a consequence,
particle spectra at small transverse momentum should be similar in nucleus-nucleus collisions, independently of the mass number, when compared at similar values of $\Npart$~\cite{Alver:2005nb}.

At high \pt, typically above 10~\gevc, particles originate from parton fragmentation and are sensitive to the amount of energy loss that the partons suffer when propagating in the medium. In a simplified model, the energy loss depends on the number of scattering centers, which is roughly proportional to the energy density, and on the path length that the parton propagates in the medium~\cite{Bjorken:1982tu}. For elastic collisions, the dependence is linear, while for medium induced gluon radiation, it is quadratic~\cite{d'Enterria:2009am}. A description of experimental data  lies in between those two~\cite{Ortiz:2017cul}.

For hard processes, the production yield \Naa in nucleus-nucleus (A--A) collisions is expected to scale with the average nuclear overlap function \avTAA when compared to
the production cross section $\sigma _{\rm pp}$ in \pp collisions. In the absence of nuclear effects, the nuclear modification factor
\begin{equation}
\raa (\pt) =  \frac{1}{\avTAA} \cdot \frac{{\rm d}\Naa (\pt)/ {\rm d}\pt}{{\rm d} \sigma _{\rm pp}(\pt) / {\rm d}\pt} 
\end{equation}
equals unity.
The average nuclear overlap function is defined as the average number of binary nucleon-nucleon collisions \avNcoll per inelastic nucleon-nucleon cross section and is estimated via a Glauber model calculation~\cite{Loizides:2017ack}.  
At the Large Hadron Collider (LHC), particle production is observed to be strongly suppressed in \pbpb collisions 
by a factor of up to 7--8 around \pt~=~6--7~\gevc with a linear decrease of the suppression factor at higher \pt but still a substantial suppression even above 100~\gevc~\cite{Acharya:2018qsh, Khachatryan:2016odn}. 

The LHC produced for the first time collisions of xenon nuclei at a center-of-mass energy of $\sqrsn~=~5.44~\tev$ during a pilot run with 6~hours of stable beams in October 2017. This allows for studying the dependence of particle production on the collision system size where
xenon neatly bridges the gap between data from \pp, \pPb and \pbpb collisions. Here, the atomic mass numbers are $A=129$ for xenon, and $A=208$ for lead 
with half-density radii of the nuclear-charge distribution of $r$~=~(5.36~$\pm$~0.1)~fm and (6.62~$\pm$~0.06)~fm, respectively~\cite{Loizides:2017ack,DEVRIES1987495}. 
The parameters of the nuclear-charge density distribution for $^{129}$Xe are not yet measured but were extrapolated from neighboring isotopes and are thus less precisely known than for $^{208}$Pb. While $^{208}$Pb is a spherical nucleus, $^{129}$Xe has a deformation parameter of $\beta_2 = (0.18 \ \pm \ 0.02)$. 

This article reports transverse momentum spectra of charged particles at mid-pseudorapidity in \xexe collisions at $\sqrsn~=~5.44~\tev$ 
measured with the ALICE apparatus at the LHC in the kinematic range $0.15 < \pt < 50\ \gevc$ and $|\eta| < 0.8$ for nine classes of collision centrality, covering the most central 80\% of the hadronic cross section. It is organized as follows:
Section~2 describes the experimental setup and data analysis. Systematic uncertainties are discussed in Sect.~3. Results and comparison to model calculations are presented in Sect.~4. A summary is given in Sect.~5.

\section{Experiment and data analysis}
Collisions of xenon nuclei were recorded at an average instantaneous luminosity of about \mbox{$2\cdot10^{-25}\ \text{cm}^{-2} \text{s}^{-1}$} and a hadronic interaction rate of 80--150~s$^{-1}$.
A detailed description of the ALICE experimental apparatus can be found elsewhere~\cite{Aamodt:2008zz}.

\subsection{Trigger and event selection}
A minimum-bias interaction trigger was optimized for high efficiency on hadronic collisions. 
It required signals from both forward scintillator arrays covering $2.8<\eta<5.1$~(V0A) and $-3.7<\eta<-1.7$~(V0C).
Additionally, coincidence with signals from two neutron Zero-Degree
Calorimeters (ZDC), ZNA and ZNC, at $|\eta| > 8.7$ was required in order to remove contamination from electromagnetic processes.
Here A and C denote opposite sides of the experiment along the beamline. 
The offline event selection was optimized to reject beam-induced background.
Background events were efficiently rejected by exploiting the timing signals in the two V0 detectors.
Parasitic collisions 
are removed by using the correlation between the sum and the
difference in arrival times as measured in each of the neutron ZDCs.
In total, $1.1 \cdot 10^{6}$ minimum-bias collisions pass the event selection and were further analyzed.

This analysis is based on tracking information from the Inner Tracking
System (ITS)~\cite{Aamodt:2010aa} and the Time Projection Chamber (TPC)~\cite{Alme:2010ke} which are located in
the central barrel of ALICE. A solenoidal magnet provides momentum dispersion in the direction transverse to the beam axis.
The nominal field strength in the ALICE central barrel is 0.5~T. However, in order to extend particle tracking and identification to the lowest possible momenta, it was reduced to 0.2~T in \xexe collisions.

The ITS is comprised of six cylindrical layers of silicon detectors with radii between 3.9 and 43.0~cm. The two innermost layers, with average radii of 3.9 cm and 7.6~cm, are equipped with Silicon Pixel Detectors (SPD); the two intermediate layers, with average radii of 15.0~cm and 23.9~cm, are equipped with Silicon Drift Detectors (SDD) and the two outermost layers,
with average radii of 38.0 cm and 43.0~cm, are equipped with double-sided Silicon Strip Detectors (SSD).
The large cylindrical TPC has an active radial range from about 85 to 250~cm and an overall length along the beam direction of 500~cm. It covers the full azimuth in the pseudo-rapidity range 
$|\eta| < 0.9$ and provides track reconstruction with up to 159 points along the trajectory of a charged particle as well as particle identification via the measurement of specific energy loss ${\rm d}E/{\rm d}x$.

The collision vertex is determined using reconstructed particle trajectories in the TPC including hits in the ITS. 
All collisions with a reconstructed vertex position 
within $\pm10\ \text{cm}$ along the beam direction from the nominal interaction point are accepted.
The collision  centrality is defined as the percentile of the hadronic
cross section corresponding to the measured charged particle multiplicity.
The centrality determination is based on the sum of the amplitudes of the V0A and V0C signals
\cite{ALICEpubCent,xexe-cent}.
Averaged quantities characterizing a centrality class such as the number of
participants~\Npart, the number of binary collisions~\Ncoll, and the nuclear overlap function~\Taa are calculated
as the average over all events in this class by fitting the experimental distribution with a Glauber Monte Carlo model that employs negative binomial distributions to model multiplicity production~\cite{ALICEpubCent,xexe-cent} (see Table~\ref{tab:Ncoll}).
The analysis is restricted to the 0--80\%
centrality range in order to ensure that effects of trigger inefficiency and
contamination by electromagnetic processes are negligible. 
\begin{table}[ht]
\centering
\begin{tabular}{@{} c| cccc| c @{}} 
\hline\hline
Centrality (\%) & $\avg{\dnchdeta}_{\textrm{Xe--Xe}}$ & \avNpart & \avNcoll  & \avTAA $\left({\rm mb}^{-1} \right)$ & $\avg{\dnchdeta}_{\textrm{Pb--Pb}}$ \\
\hline
0--5       & 1167 $\pm$ 26 & 236  $\pm$ 2    & 949 $\pm$53 & 13.9 $\pm$0.8  &  1943 $\pm$ 54\\
5--10     &  939 $\pm$ 24 & 207  $\pm$ 2    & 737 $\pm$46 & 10.8 $\pm$0.7  &  1586 $\pm$ 46\\
10--20   &  706 $\pm$ 17 & 165  $\pm$ 2   & 511 $\pm$26  & 7.5 $\pm$0.5  &  1180 $\pm$ 31\\
20--30   &  478 $\pm$ 11 & 118  $\pm$ 3    & 303 $\pm$28   & 4.4 $\pm$0.4  &  786 $\pm$ 20\\
30--40   &  315 $\pm$   8 & 82  $\pm$ 3    & 171 $\pm$19   & 2.5 $\pm$0.3  &  512 $\pm$ 15\\
40--50   &  198 $\pm$   5 & 55 $\pm$ 3  & 92 $\pm$11   & 1.3 $\pm$0.2  &  318 $\pm$ 12\\
50--60   &  118  $\pm$  3 & 34 $\pm$ 2  & 46 $\pm$6 & 0.7 $\pm$0.1  &  183 $\pm$ 8\\
60--70   &    65 $\pm$  2 & 20 $\pm$ 2 & 22 $\pm 3$ & 0.32 $\pm$0.04  &  96 $\pm$ 6\\
70--80   &    32 $\pm$  1& 11 $\pm$ 1  & 10 $\pm $1 & 0.14 $\pm$0.02  &  $45 \pm$ 3\\
\hline\hline
\end{tabular}
\caption{\label{tab:Ncoll} Averaged values of \avg{\dnchdeta}, \avg{\Npart}, \avg{\Ncoll} and \avg{\Taa}
  for nine centrality classes of \xexe collisions~\cite{ALICEpubCent,xexe-cent} at $\sqrsn$~=~5.44~\tev, and \avg{\dnchdeta} for \pbpb collisions at $\sqrsn$~=~5.02~\tev~\cite{Adam:2015ptt}.
  The values for \avg{\dnchdeta} are measured in the range $|\eta| < 0.5$.}
\end{table}

\subsection{Track selection}
\label{sec:ana}
Primary charged particles within the kinematic range $|\eta|<0.8$ and $0.15 < \pt < 50$~GeV/$c$ are measured.  
Here, primaries are defined as all charged particles with a proper lifetime $\tau$ larger than 1~cm/$c$ that are either produced directly in the primary beam-beam interaction, or from decays of particles with $\tau$ smaller than 1~cm/$c$, excluding particles produced in interactions with the detector material~\cite{Acharya:2017dpp}.
The track selection is optimized for best track quality and minimum contamination from secondary particles. The selection criteria are identical to those of the previous analysis of \pbpb collisions at $\sqrsn$ = 5.02~\tev~\cite{Acharya:2018qsh} except for the following changes in the parameterization on the transverse momentum dependence.
The geometrical track length in the TPC fiducial volume~\cite{Alme:2010ke} is $L/{\rm (1\, cm)} > 130 - \left( \pt / (1\, \gevc) \right ) ^{-0.7}$, and
the distance of closest approach to the primary vertex in the transverse plane is $|{\rm DCA}_{{\rm xy}}| /{\rm (1\, cm)} < 0.0119+ 0.049 \left( \pt / (1\, \gevc) \right)^{-1}$. These changes reflect differences in particle tracking due to the reduced magnetic field.
In order to reject fake tracks that contaminate the spectrum, especially at high \pt, another selection is introduced: the uncertainty in the reconstructed \pt as estimated from the covariance matrix of the track fit must be less than ten times the standard deviation, when averaged over all tracks at that momentum.

\subsection{Corrections}
The doubly-differential transverse momentum spectra in \xexe collisions are normalized by the number of events $N_{\textrm{ev}}$ in each centrality class, and are given by 
\begin{equation} \label{eq:inv_yields}
 \frac{1}{N_{{\rm ev}}}  \frac{{\rm d}^{2} N_{\rm{ch}}} {{\rm d}\eta {\rm d}\pt}
  \equiv  \frac{N_{\rm{ch}}^{\rm {rec}}(\Delta\eta, \Delta \pt) }{N_{\rm {ev}} \cdot \Delta\eta \Delta\pt} \cdot \frac{\delta_{\pt}(\Delta \pt)}{\alpha (\Delta \pt) \cdot \epsilon(\Delta \pt)}, 
\end{equation}
where $N_{\rm{ch}}^{\rm {rec}}$ is the raw yield of reconstructed primary charged particles in each interval of pseudo-rapidity and transverse momentum $(\Delta\eta,\Delta \pt)$.
The symbols $\alpha(\Delta \pt)$ and $\epsilon(\Delta \pt)$ are the correction factors for detector acceptance and tracking efficiency, respectively. The correction due to the finite transverse-momentum resolution 
in the reconstruction of primary charged particles is denoted by $\delta_{\pt}(\Delta \pt)$. 
The efficiencies for trigger, event vertex reconstruction and tracking are estimated using Monte Carlo simulations with HIJING \cite{Wang:1991hta} as the event generator and GEANT3~\cite{GEANT3} for particle propagation and simulation of the detector response. The trigger and vertex selections are fully efficient for the whole centrality range used in the analysis. 

\begin{figure}[!hbt]
	\centering
	\includegraphics[width=0.6\textwidth]{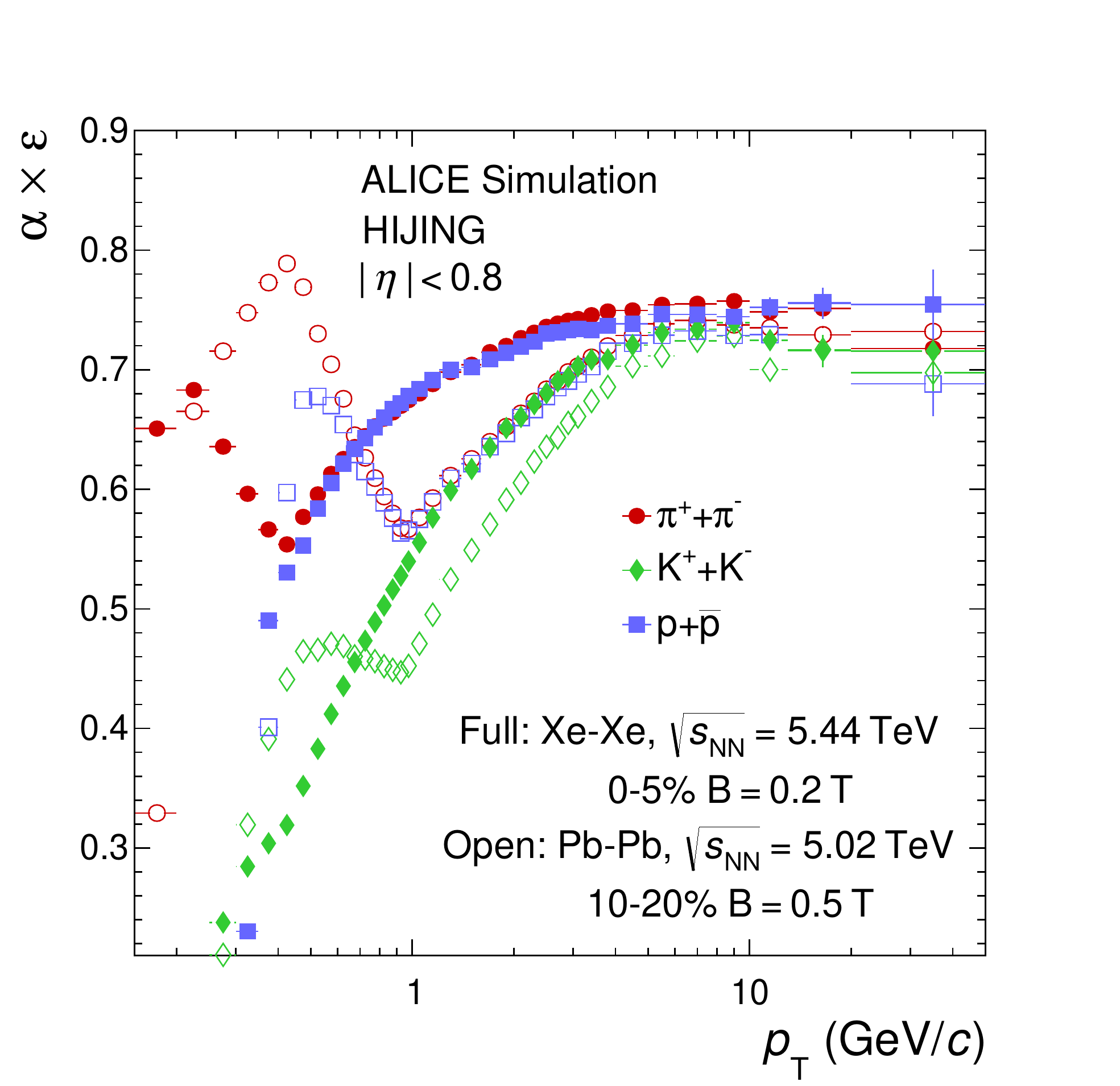}
	\caption{Transverse momentum dependence of the acceptance times tracking efficiency for the 5\% most central \xexe collisions and comparison to the 10--20\% centrality class for \pbpb collisions. The two centrality classes have similar multiplicity densities.}
	\label{fig:corr}
\end{figure}

Contamination from secondary charged particles, i.e. from weak decays and interactions in the detector material, is subtracted from the raw spectrum by employing a data driven method~\cite{Acharya:2018qsh}.
Reconstructed trajectories of primary charged particles point to the collision vertex, while charged particles from weak decays and particles generated in the detector material preferentially point away from it. 
In order to distinguish between primary and secondary particles, 
the distance of closest approach to the collision vertex in radial direction, ${\rm DCA}_{{\rm xy}}$, is used. 
A multi-template function that consists of templates for primary particles, secondary particles produced from weak decays and secondary particles from interactions in the detector material is fitted to the ${\rm DCA}_{{\rm xy}}$ distributions in each \pt interval. 

The primary charged particle reconstruction efficiency is obtained from the Monte Carlo simulation. As discussed in detail
in~\cite{Acharya:2018qsh}, this efficiency depends on the relative abundances of the various primary particles species. These relative
abundances are adjusted in the simulation using a data-driven re-weighting procedure.
The particle composition in \xexe collisions is not yet known. However, bulk particle production scales with the average charged particle multiplicity density, \avg{\dnchdeta}, independently of the collision system~\cite{ALICE:2017jyt}. In \xexe collisions, the weights from existing analyses~\cite{Abelev:2013vea,Abelev:2013xaa,Abelev:2014laa,Adam:2016tre,Acharya:2018qsh}
with \pbpb collisions at $\sqrsn$ = 2.76~\tev at equivalent values in \avg{\dnchdeta} are applied.

The acceptance times tracking efficiency for charged pions, charged kaons and (anti-)protons for \mbox{5\% most} central \xexe collisions is shown in Fig.~\ref{fig:corr}
 as a function of the particle transverse momentum
 and compared to 10--20\% \pbpb collisions at $\sqrsn$ = 5.02~\tev. The two centrality classes have similar multiplicity densities.
The particular shape with a dip at $\pt\sim 0.4~\gevc$ arises from the geometrical length selection that is especially visible for pions. This dip corresponds to particles that cross the TPC sector boundaries under small angles.
The decrease at low values of \pt is due to curling trajectories in the magnetic field which do not reach the required minimum track length in the TPC and due to energy loss and absorption in the detector material.
In \pbpb collisions, the magnetic field was set to $B=0.5$~T, which results in the dip being positioned around 1~\gevc. 
At large \pt, above 7~\gevc, the tracking efficiency is reduced by an increased local track density, i.e. high \pt particles are preferentially produced within jets, leading to a slight decrease in the track finding performance.

The transverse momentum of primary charged particles is reconstructed from the
track curvature as measured by the ITS and the TPC~\cite{Abelev:2014ffa}.  
The finite momentum resolution modifies the reconstructed charged-particle spectrum and is estimated by the 
corresponding covariance matrix element of the Kalman fit.
The relative \pt resolution, $\sigma(\pt)/\pt$, depends on the momentum and amounts to approximately 4.5\% at $\pt = 0.15$~\gevc,
it shows a minimum of 1.5\% around $\pt=1.0$~\gevc, and increases linearly for
larger \pt, approaching 9.3\% at 50~\gevc.
The centrality dependence of the relative \pt resolution is negligible.
To account for the finite \pt resolution, correction factors to the spectra are determined
from an unfolding procedure as described in ~\cite{Acharya:2018qsh}, using Bayesian unfolding at low \pt and a bin-by-bin correction at large \pt. 
The \pt dependent correction factors are
applied to the measured \pt spectrum and depend slightly on collision centrality because of the change in the slope of the spectrum at high \pt. 
At transverse momenta below 10~\gevc, $\delta_{\pt}$ deviates significantly from unity only at the lowest momentum interval of $0.15 \le \pt < 0.2$~\gevc where it amounts to 0.5\% for all centrality classes, and by up to 3\% (4\%) in 0--5\% (70--80\%) central collisions above 10~\gevc.

The statistical uncertainty of the spectra is dominated by the statistical uncertainty in the raw data. It is largest at the highest momentum interval of 40-50~\gevc and amounts to 28\% (38\%) for the 0--5\% (30--40\%) centrality class while the contribution from the Monte Carlo efficiency is 2\% (4\%) or less. 

\subsection{\pp reference at $\mathbf{\s}$ = 5.44~\text{TeV}}
\begin{figure}[hbt]
	\centering
	\includegraphics[width=0.6\textwidth]{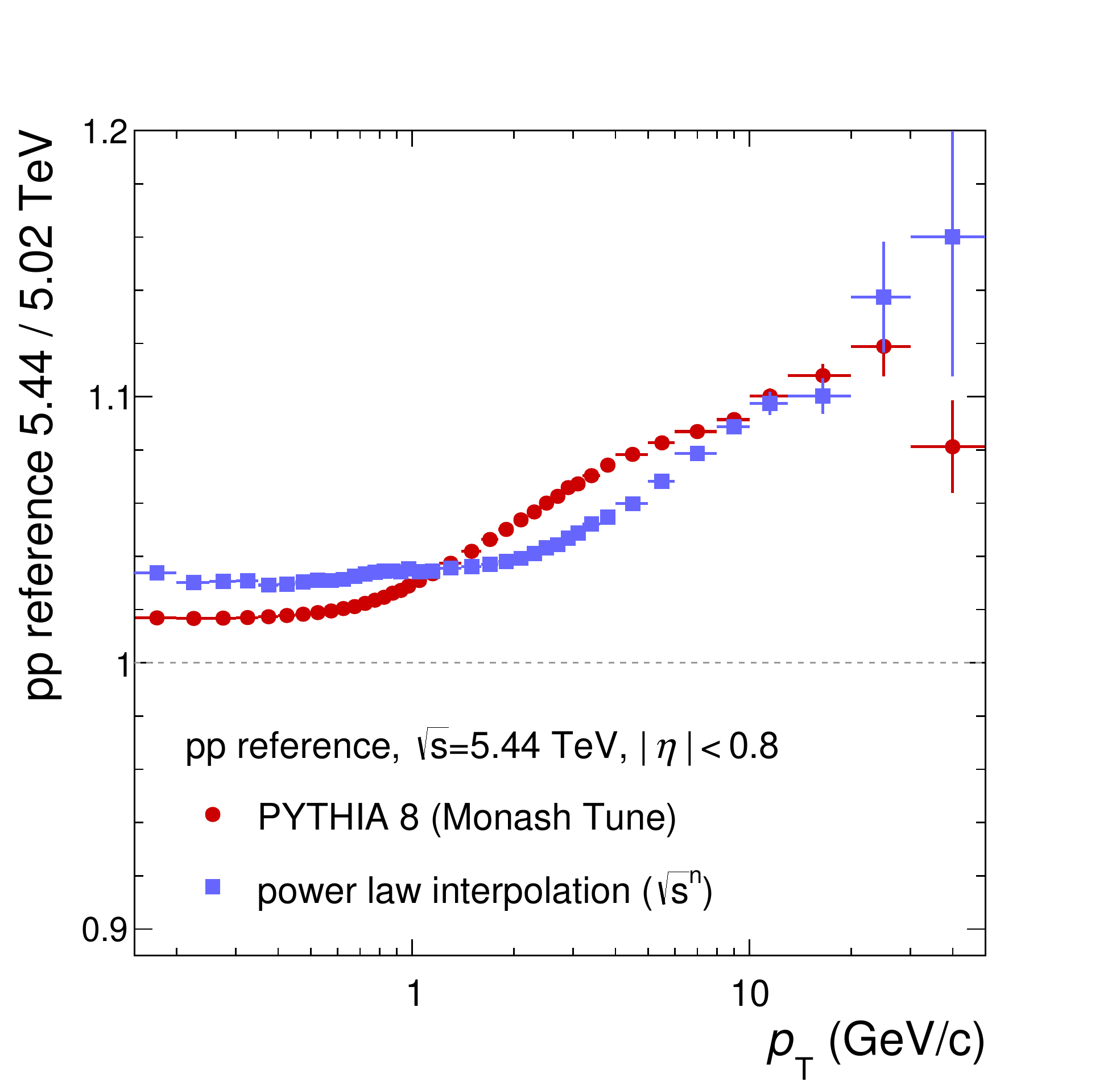}
	\caption{Ratio of \pt-differential inelastic cross sections in \pp collisions at $\sqrs$~=~5.44~\tev over 5.02~\tev using a power law interpolation and the event generator PYTHIA~8.}	
	\label{fig:pp-ref}
\end{figure}

The \pt-differential inelastic cross section in \pp collisions at $\s = 5.44~\tev$ is needed to measure the corresponding nuclear modification factor. 
As there are no measurements of \pp collisions at this energy, a reference is obtained by interpolating
\pp references as measured at $\sqrs$~=~5.02~\tev and $\sqrs$~=~7~\tev assuming a power-law dependence in each \pt interval, $d\sigma / d \pt (\sqrs)  \propto  \sqrs^{n}$.
The value of the free parameter $n$ varies between $0.35$ and 1.75, depending on \pt. 
This approach is a combination of the interpolation method that was used over the full \pt range in~\cite{Adam:2016dau} and for $\pt < 5$~\gevc as used in~\cite{Abelev:2013ala}.
The statistical uncertainty of the \pp reference is interpolated between the references at $\sqrs$~=~5.02~\tev and 7~\tev assuming also a power-law dependence and is assigned to the interpolated reference.
It amounts to 7.8\% at the momentum interval of 30-50~\gevc.  

As an alternative approach, the scaling of the measured cross section at $\sqrs$ = 5.02~\tev to  $\sqrs$~=~5.44~\tev by using the ratio of spectra at those two energies obtained with the PYTHIA~8 (Monash tune) event generator~\cite{Skands:2014mon} is studied.
The ratio of the \pp references at $\sqrs$~=~5.44~\tev from the power-law interpolation and at $\sqrs$ = 5.02~\tev  is shown in Fig.~\ref{fig:pp-ref} together with
results obtained with the alternative method.
The spectrum is harder at higher collision energy, with a small change in the total cross section of 4\% below 1~\gevc and an increase of about 10\% at transverse momenta above 10~\gevc.

\section{Systematic uncertainties}
For the total systematic uncertainty, all contributions are added in quadrature and are summarized in Table~\ref{tab:CombinedSyst}.
\begin{table} [hpt]
	\centering 
    \setlength\extrarowheight{2pt}
    \begin{tabular}{lccc}
    	\hline\hline
         centrality  (\%)       	& 0--5 (\%) & 30--40 (\%) & 70--80 (\%)  \\ [0.5ex]
        \pt range (\gevc)               &  0.2--0.5  /  1--2  /  40--50 &  0.2--0.5  /  1--2  /  40--50 &  0.2--0.5  /  1--2  /  40--50\\ 
        \hline 	
        Source\\
        Vertex selection      			& 0.2 / 0.2 / 0.2					& 0.8 / 0.8 / 0.8 	& 0.8 / 0.8 / 0.8\\
        Track selection                	& 1.6 / 0.9 / 1.2 		& 0.9 / 0.6 / 0.8 		& 0.9 / 0.5 /  1.0\\
        Secondary particles             & 1.4 / 0.2 / negl. 	& 0.8 / 0.2 / negl. 	& 0.6 / 0.2 / negl. \\
        Particle composition            & 0.3 / 1.7 / 0.7 		& 0.4 / 1.9 / 1.0 		& 0.7 / 0.6 / 0.6\\
        Tracking efficiency             & 1.9 / 1.2 / 0.4		& 2.2 / 1.2 / 0.4 		& 2.2 / 1.4 / 0.6\\
        Material budget                 & 0.3 / 0.3 / 0.1		& 0.3 / 0.3 / 0.1		& 0.3 / 0.3 / 0.1\\
        \pt resolution                  & negl. / negl. / 0.5	& negl. / negl. / 0.7	& negl. / negl. / 0.9\\
        \hline
        Sum, \pt dependent:             & 3.1 / 2.4 / 1.5		& 2.8 / 2.5 / 1.8		& 2.8 / 1.9 / 2.1 \\
        \hline
        Centrality selection			& 0.1					& 0.8					& 3.2\\ 
        \hline\hline
    \end{tabular}
    \caption{Contributions to the systematic uncertainty in units of percent for the 0--5\%, 30--40\%, and 70--80\% centrality classes in \xexe collisions. The numbers are averaged in the \pt intervals from 0.2--0.5~\gevc (left), 1--2~\gevc (middle) and 40--50~\gevc (right). For the \pt-dependent sum, contributions are added in quadrature.}
    \label{tab:CombinedSyst}
\end{table}

The effect of the selection of events based on the vertex position is studied by comparing the fully corrected \pt spectra obtained 
with alternative vertex selections corresponding to $\pm$~5~cm,  and $\pm$~20~cm. 
The difference in the fully corrected \pt spectra is less than 0.3\% for central collisions and less than 0.5\% for peripheral collisions.

In order to test the description of the detector response and the track reconstruction in the simulation, all criteria for track selection are varied within the ranges as described in the previous publication~\cite{Acharya:2018qsh}. A full analysis is performed by varying one selection criterion at a time.
The maximum change in the corrected \pt spectrum is then considered as systematic uncertainty.
The overall systematic uncertainty related to track selection is obtained from summing up all individual contributions quadratically and it amounts to 0.6--3.0\%, depending on \pt and centrality.

The systematic uncertainty on the secondary-particle contamination is estimated by varying the fit model using two templates, i.e.~for primaries and secondaries, or
three templates, i.e.~primaries, secondaries from interactions in the detector material and secondaries from weak
decays of K$^{0}_{{\rm s}}$ and $\Lambda$, as well as varying the
fit ranges. The maximum difference between data and the 
two-component-template fit is summed in quadrature together with the difference
between results obtained from the two- and three-component-template fits. 
The systematic uncertainty due to the contamination from secondaries is decreasing with increasing \pt. It dominates at low \pt with values up to 4\% and is negligible above 2~\gevc.

The systematic uncertainty on the primary particle composition is taken from~\cite{Acharya:2018qsh}. An additional uncertainty is estimated by assuming the particle composition from a neighboring \avg{\dnchdeta} range to the matched one in the \pbpb analysis and is added quadratically. 
The sum peaks around 3~\gevc with a maximum of 5\% (less than 2\%) for the 0-5\% (70--80\%) centrality class.

In order to estimate the systematic uncertainty due to the tracking efficiency, the track matching between the TPC and the ITS information in data and Monte Carlo is compared after scaling the fraction of secondary particles obtained from the fits to the ${\rm DCA}_{{\rm xy}}$ distributions~\cite{Acharya:2018qsh}. 
The difference in the TPC-ITS track-matching efficiency between data and simulation 
is assigned to the corresponding systematic uncertainty (see Table~2).
It amounts to 2\% in central collisions, and up to 3.5\% in peripheral collisions.

The material budget in ALICE at $\eta \approx 0$ amounts to $(11.4\ \pm \ $0.5)\% in radiation lengths for primary charged particles that have sufficient track length in the TPC~\cite{Abelev:2014ffa}.
A difference in the amount of detector material leads to different amounts of secondary particles that are produced. 
After the subtraction of the contribution due to secondaries using the three-component ${\rm DCA}_{{\rm xy}}$ fits, the differences on the secondary correction factor is negligible.
A variation of the material budget within above limits leads to a \pt dependent systematic uncertainty on the tracking efficiency of 0.1--0.3\%.

The uncertainty due to the finite \pt resolution at high \pt is estimated using the azimuthal
dependence of the 1/\pt spectra for positively and negatively charged particles. The relative shift of
the spectra for oppositely charged particles along 1/\pt  determines the size of uncertainty for a
given angle. The RMS of the 1/\pt shift as distributed over the full azimuth is used as an additional increase of the \pt resolution.
For the lowest \pt bin the uncertainty is estimated from the unfolding procedure applied to Monte Carlo simulations.
The uncertainty due to the finite \pt resolution is significant only at the lowest and highest momenta bins and amounts to 0.5\% at the lowst \pt bin for all centralities and 0.5\% (0.9\%) for the 0-5\% (70--80\%) centrality class.

The uncertainty due to the centrality determination is estimated by changing the fraction of the visible cross section $(90.0\ \pm \ 0.5)$\%.
The uncertainty is estimated from the variation of the resulting \pt spectra and amounts to $\sim$~0.1\% and $\sim$~3.2\% for central (0--5\%) and peripheral (70--80\%) collisions, respectively.

The systematic uncertainty of the \pp reference at \sqrs~=~5.44~\tev has two contributions, which are added quadratically.
For each \pt interval, the systematic uncertainty of the \pp references at \sqrs~=~5.02~\tev and  \sqrs~= 7~\tev are interpolated to \sqrs~=~5.44~\tev by using a power-law.
This corresponds to interpolating between the upper and lower boundaries of the experimental data points as given by their systematic uncertainties. 
It assumes full correlation of systematic uncertainties at both energies.

The difference between the interpolated reference and the one using the PYTHIA~8 event generator is assigned as the other contribution 
to the systematic uncertainty in the pp reference, in each \pt interval. The systematic uncertainty in the \pp reference has a minimum of 2.2\% around 1~\gevc and reaches its maximum of 7.7\% at the highest momentum bin.

\section{Results}
\label{sec:results}
\begin{figure}[H]
	\center
	\includegraphics[width=0.60\textwidth]{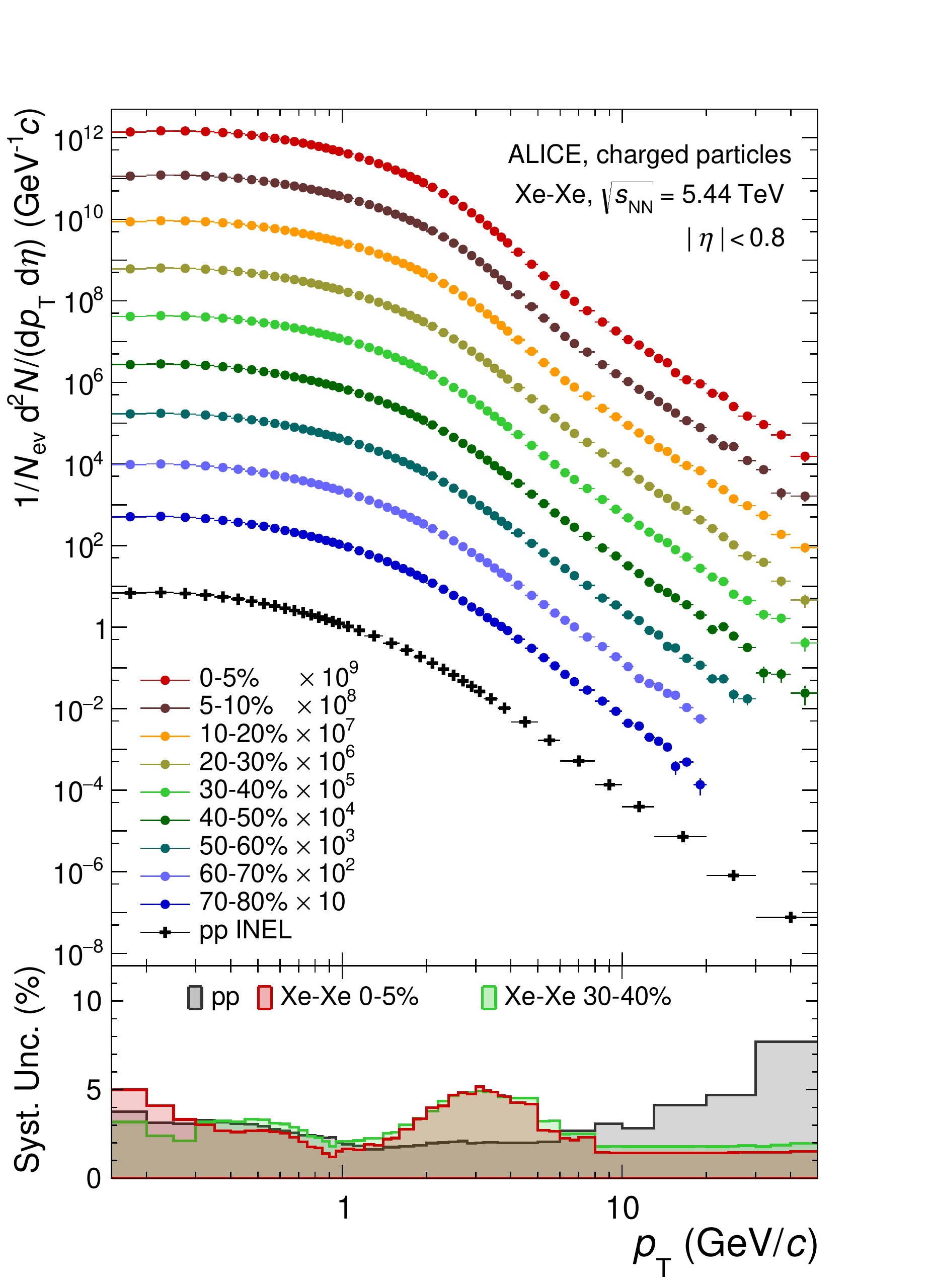}
	\caption{Transverse momentum spectra of charged particles in \xexe collisions at \sqrsn~=~5.44~\tev in nine centrality classes together with the
interpolated \pp reference spectrum at \sqrs~=~5.44~\tev (top panel) and systematic uncertainties (bottom panel).}
	\label{fig:finalSpectra}
\end{figure}

The transverse momentum spectra of charged
particles in \xexe collisions 
are shown in the top panel of Fig.~\ref{fig:finalSpectra} for nine centrality classes together with the
interpolated \pp reference spectrum at $\sqrs$~=~5.44~\tev. The latter is obtained from the interpolated \pt-differential cross section by dividing it by the interpolated inelastic nucleon-nucleon cross section of $(68.4\ \pm \ 0.5)$~mb at $\sqrs$~=~5.44~\tev~\cite{Loizides:2017ack}. 
In the most-peripheral collisions, the \pt spectrum is similar to that of \pp collisions and exhibits a power law
behavior that is characteristic of  hard-parton scattering and vacuum
fragmentation. With increasing collision centrality, the \pt differential cross section is progressively depleted above 5~\gevc.

Systematic uncertainties are shown in the bottom panel. At momenta between 0.4 and 10~\gevc, the systematic uncertainty is dominated by the contribution from tracking and amounts to about 2--3\%. 
It is almost independent of \pt above 10~\gevc with a value of 1.4\% (2.1\%) for the 0--5\% (70--80~\%) centrality class.  

\begin{figure}[hbt]
	\center
	\includegraphics[width=0.9\textwidth]{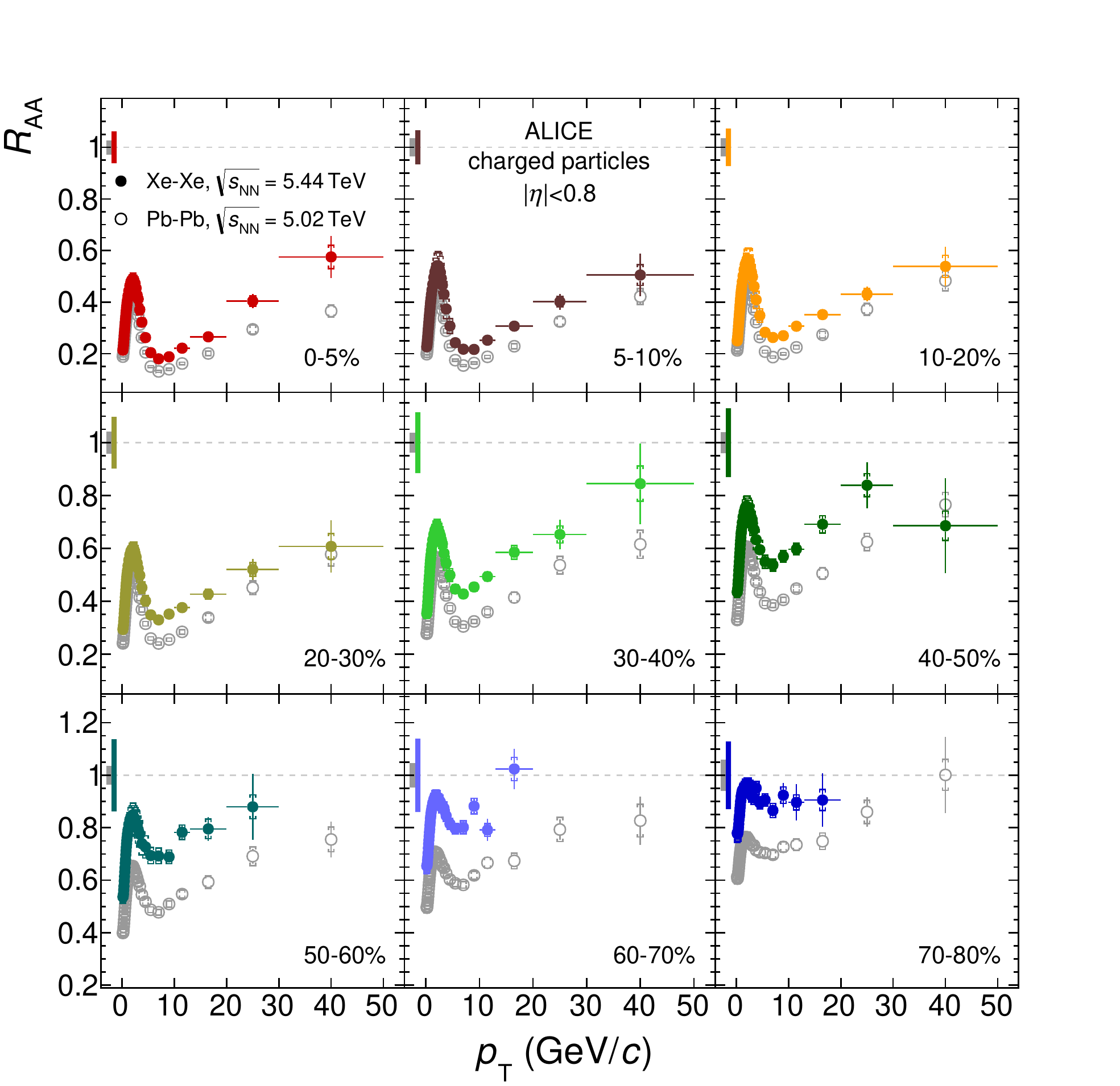}
	\caption{Nuclear modification factor in \xexe at \sqrsn~=~5.44~\tev (filled circles) and \pbpb collisions~\cite{Acharya:2018qsh} at \sqrsn~=~5.02~\tev (open circles) for nine centrality classes. The vertical lines (brackets) represent the statistical (systematic) uncertainties. The overall normalization uncertainty is shown as a filled box around unity.}
	\label{fig:finalRxexe}
\end{figure}

In order to determine the nuclear modification factor \raa, the interpolated \pt-differential \pp cross section is scaled by the average nuclear overlap function~\avTAA. 
The resulting nuclear modification factor as a function of transverse momentum is shown in Fig.~\ref{fig:finalRxexe} for nine centrality classes and compared to results from \pbpb collisions~\cite{Acharya:2018qsh}.
The overall normalization uncertainties for \raa are indicated by vertical bars around unity. 
The uncertainties of the \pp reference and the centrality determination are added in quadrature.
The latter is larger for \xexe collisions than for \pbpb because of the less precisely known
nuclear-charge-density distribution of the deformed $^{129}$Xe and the resulting larger relative uncertainty in \avTAA~\cite{ALICEpubCent,xexe-cent}.
The nuclear modification factor exhibits a strong centrality
dependence with a minimum around \mbox{$\pt=6$--7~\gevc} and an almost
linear rise above. In particular, in the 5\% most central \xexe collisions, at the minimum, the yield is suppressed by a
factor of about 6 with respect to the scaled \pp reference. The nuclear modification factor reaches a value of 0.6 at the highest measured transverse-momentum interval of 30--50~\gevc.
For comparison, the nuclear modification factor \raa in \pbpb collisions at \sqrsn~=~5.02~\tev is shown in Fig.~\ref{fig:finalRxexe} as open circles for the same centrality classes as \xexe.
In both collision systems, a similar characteristic \pt dependence of \raa is observed. In \pbpb collisions, the suppression of high-momentum particles is apparently stronger for the same centrality class but still in agreement with \xexe collisions within uncertainties.

\begin{figure}[hbt]
	\center
	\includegraphics[width=0.75\textwidth]{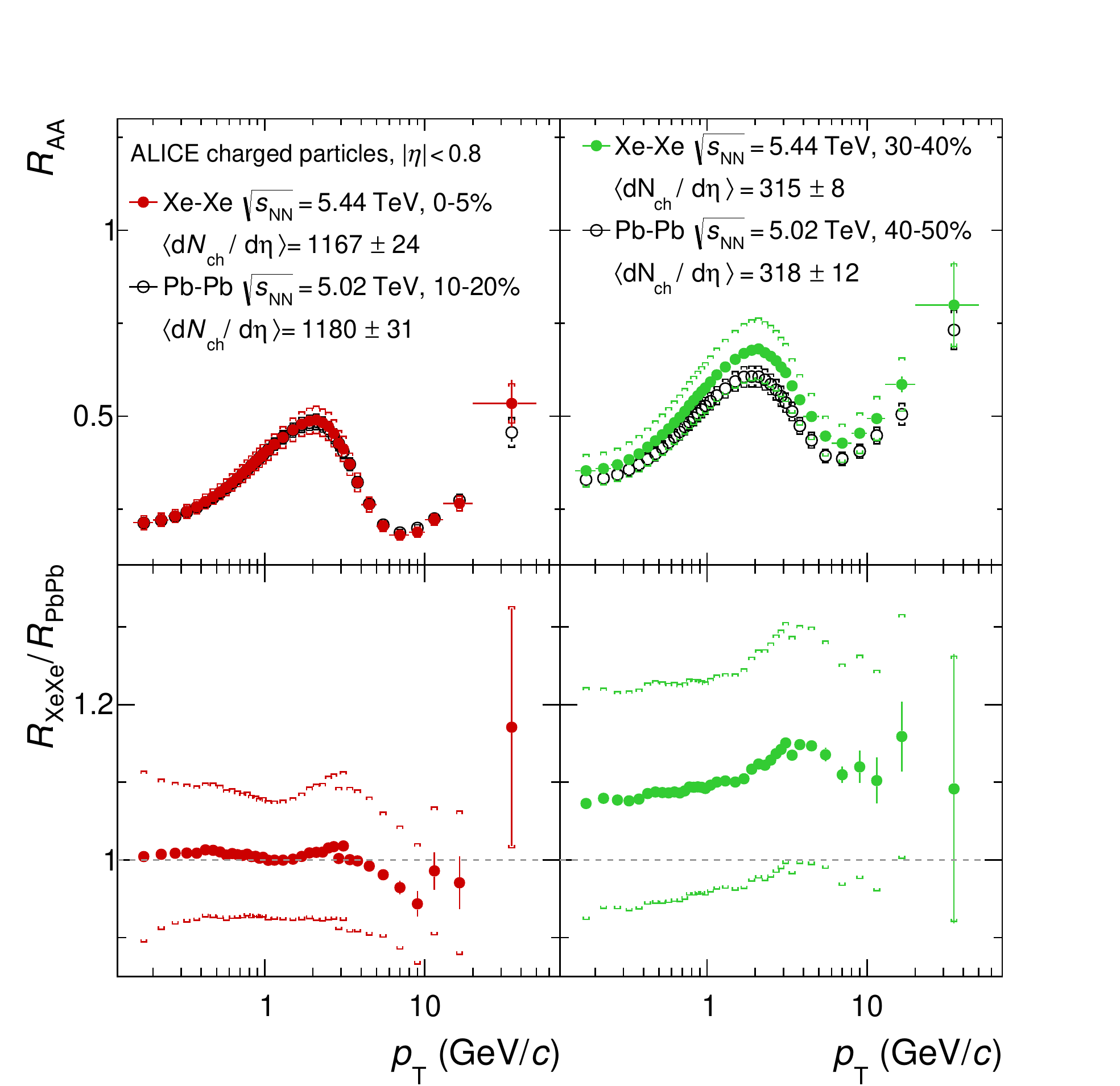}
	\caption{Comparison of nuclear modification factors in \xexe collisions (filled circles) and \pbpb collisions (open circles) for similar ranges in \avg{\dnchdeta} for the 0--5\% (left) and 30--40\% (right) \xexe centrality classes. The vertical lines (brackets) represent the statistical (systematic) uncertainties.}
	\label{fig:raa_PbPb_XeXe_dNchdeta}
\end{figure}

Nuclear modification factors from \xexe and \pbpb collisions and their ratios at similar ranges of \avg{\dnchdeta} are shown in Fig.~\ref{fig:raa_PbPb_XeXe_dNchdeta}.
In 5\% most central \xexe collisions, the nuclear modification factor is remarkably well matched by 10--20\% central \pbpb collisions over the entire \pt range. 
ln the 30--40\% \xexe (40--50\% \pbpb) centrality class, 
again agreement is found within uncertainties.
These findings of matching nuclear modification factors at similar ranges of \avg{\dnchdeta}
are in agreement with results from the study of fractional momentum loss of high-\pt partons at RHIC and LHC energies~\cite{Adare:2015cua}. 

\begin{figure}[hbt]
	\center
	\includegraphics[width=0.68\textwidth]{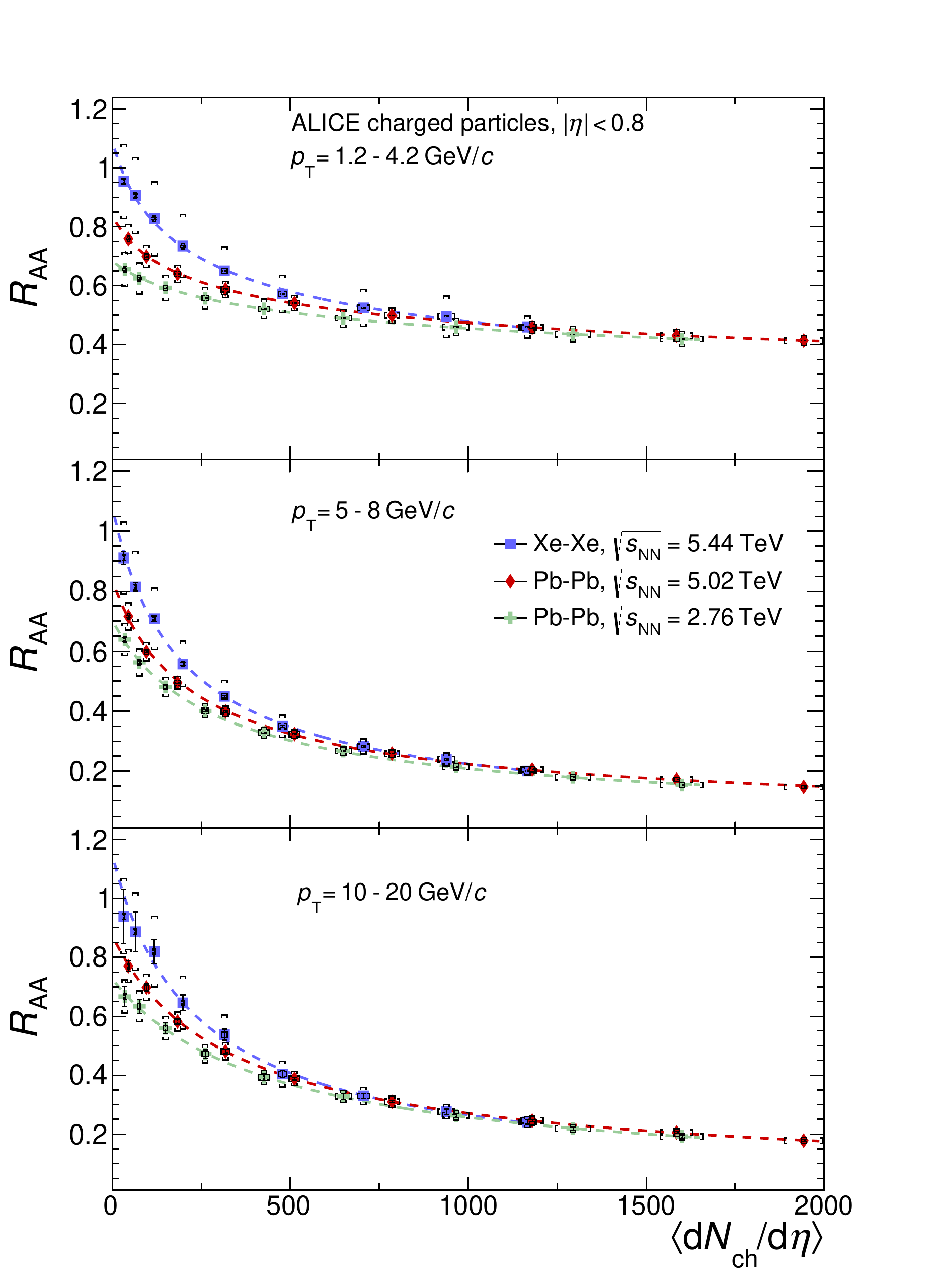}
	\caption{Comparison of the nuclear modification factor in \xexe and \pbpb collisions integrated over identical regions in \pt as a function of \avg{\dnchdeta}. The vertical brackets indicate the quadratic sum of the total systematic uncertainty in the measurement and the overall normalization uncertainty in \avTAA.The horizontal bars reflect the RMS of the distribution in each bin. The dashed lines show results from power-law fits to the data and are drawn to guide the eye.}
	\label{fig:raaVsNch}
\end{figure}

A  comparison of the nuclear modification factors as a function of \avg{\dnchdeta} in \xexe and \pbpb collisions for three different regions of \pt (low, medium, and high) is shown in Fig.~\ref{fig:raaVsNch}.
A remarkable similarity in \raa is observed between \xexe collision at \sqrsn~=~5.44~\tev and \pbpb collisions at \sqrsn~=~5.02 and 2.76~\tev when compared at identical ranges in \avg{\dnchdeta}, for $\avg{\dnchdeta} > 400$.  This holds both at low momentum where the hydrodynamical expansion of the medium dominates the spectrum and at high momentum, where parton energy loss inside the medium drives the spectral shape. 
At $\avg{\dnchdeta} < 400$, the values of \raa still agree within rather large uncertainties although no definitive conclusion can be drawn because, in particular, event selection and geometry biases could affect the spectrum in peripheral A--A collisions~\cite{Morsch:2017brb}.

In a simplified radiative energy loss scenario when assuming identical thermalization times~\cite{Giacalone:2017dud, Kolb:2000fha}, the average energy loss $\langle\Delta E\rangle$ is
proportional to the density of scattering centers in the medium, which in turn is proportional to the energy density $\epsilon$, and to the square of
the path length $L$ of the parton in the medium, $\langle\Delta E\rangle \propto \epsilon \cdot L^2$~\cite{d'Enterria:2009am}. The energy density can be estimated from the average charged-particle multiplicity density~\cite{ex:PbPb:CMSTransversEnergy:2012} per transverse area, $\epsilon \propto \avg{\dnchdeta}/\at$. In central collisions, the initial transverse area \at is related to the radius $r$ of the colliding nuclei, $\at = \pi \cdot r^2$~\cite{d'Enterria:2009am}. 
Therefore, the comparison of the measured $R_{\rm AA}$ values in the two colliding systems could enable a test of the path length dependence of medium-induced parton energy loss~\cite{Djordjevic:2018}.

To further address bulk production, the average transverse momentum \mpt in the range from 0--10~\gevc is derived. 
The spectra are extrapolated down to \pt = 0 by fitting a Hagedorn function~\cite{Hagedorn:1983tyu} in the range $0.15\, \gevc <\pt<1\gevc$.
The relative fraction of the extrapolated particle yield amounts to 8\% (11\%) for the 0--5\% (70--80\%) centrality class.
Statistical uncertainties in \mpt are negligible.
Systematic uncertainties are estimated by varying each source of systematic uncertainty in the spectra at a time, by varying the fit range to $0.15\, \gevc <\pt<0.5\gevc$, and by changing the interpolation range to 0--0.2~\gevc. 
All contributions are then added quadratically. The relative systematic uncertainty is 1.8\% (1.3\%) for the 0-5\% (70--80\%) centrality class.

\begin{figure}[H]
	\center
	\includegraphics[width=0.77\textwidth]{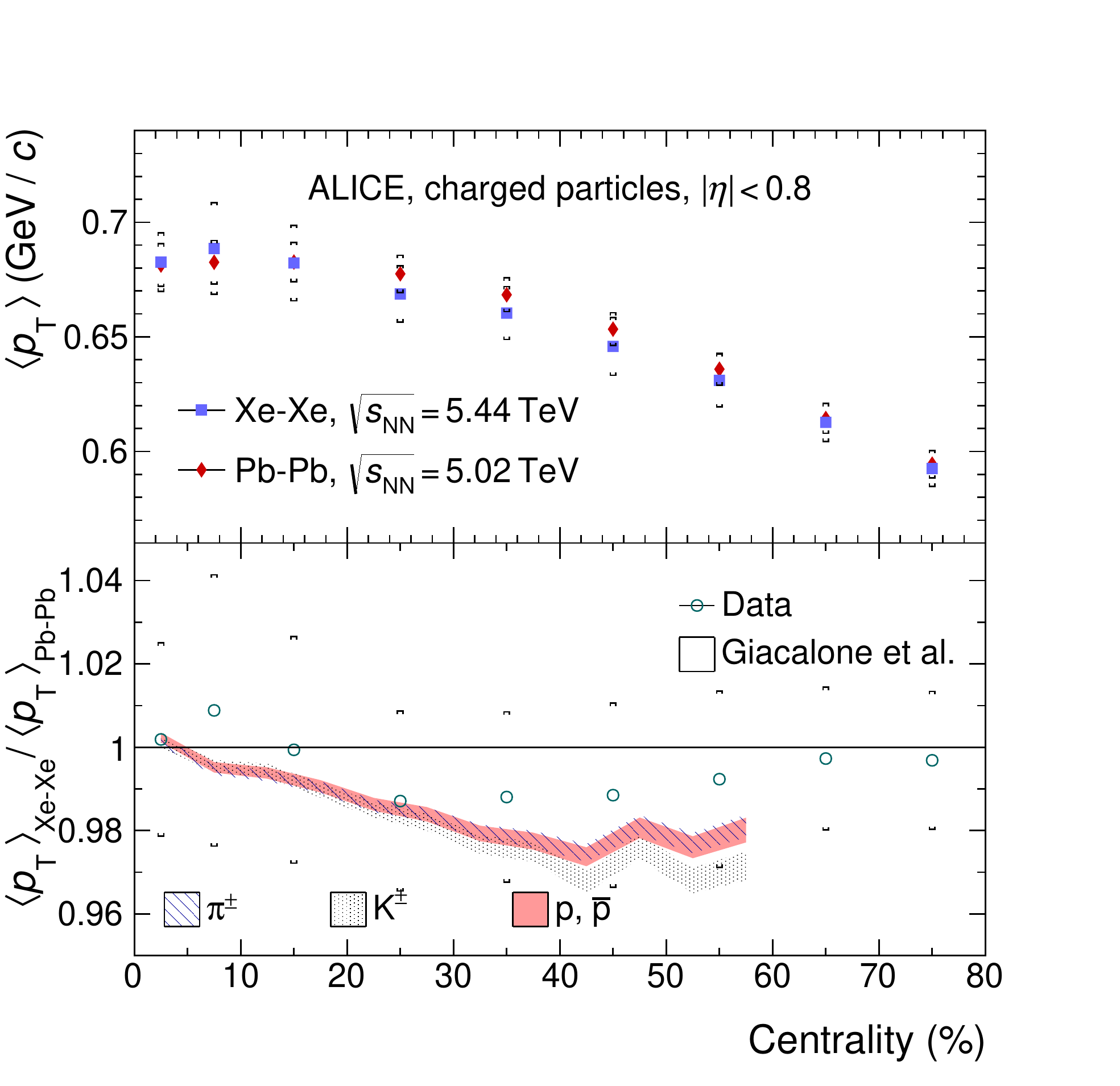}
	\caption{Average transverse momentum in the \pt-range 0--10~\gevc for \xexe collisions at \sqrs~=~5.44~\tev (squares) and \pbpb collisions at \sqrs~=~5.02~\tev (diamonds) for nine centrality classes (top) and their ratios (bottom). The vertical brackets indicate systematic uncertainties. The hashed areas show results from hydrodynamical calculations by Giacalone et al.~\cite{Giacalone:2017dud}.}
	\label{fig:mpt}
\end{figure}

The average transverse momentum is presented in the top panel of Fig.~\ref{fig:mpt} for \xexe collisions at \sqrs~=~5.44~\tev (squares) and \pbpb collisions at \sqrs~=~5.02~\tev (diamonds) for nine
centrality classes. An increase of \mpt with centrality is visible in both collision systems and is attributed to the increasing transverse radial flow. The bottom panel of  Fig.~\ref{fig:mpt} shows the ratios of \mpt in both collision systems. The ratio is flat within uncertainties but allows for relative variations of up to two percent. Comparison to results from hydrodynamical calculations~\cite{Giacalone:2017dud} are shown by the hashed areas for pions, kaons and protons.
While the calculations are not able to predict absolute particle spectra, predictions are made for the relative difference in \mpt between both collision systems in order to study the system size dependence. The predicted trend of a larger \mpt  in 5\% most central \xexe collision and continuously lower values towards the 40--50\% centrality class are consistent with the data. 

\section{Summary}
\label{sec:summary}
Transverse momentum spectra and nuclear modification factors of charged particles in \xexe collisions at $\sqrsn$~=~5.44~\tev 
in the kinematic range $0.15 < \pt < 50\ \gevc$ and $|\eta| < 0.8$ are reported for nine centrality classes, in the 0--80\% range.
A \pp reference at  $\sqrs$~=~5.44~\tev is obtained by the interpolation of the existing spectra at $\sqrs$~=~5.02 and 7~\tev.
When comparing nuclear modification factors at similar ranges of averaged charged particle multiplicity densities, 
a remarkable similarity between central \xexe collisions and \pbpb collisions at a similar center-of-mass energy of $\sqrsn$~=~5.02~\tev and at 2.76~\tev is observed for $\avg{\dnchdeta} > 400$. 
The centrality dependence of the ratio of the average transverse momentum \mpt  in \xexe collisions over \pbpb collisions is 
flat within uncertainties but allows for relative variations of up to two precent.
Predictions from hydrodynamical calculations that take into account the significantly different geometries of both collision systems are consistent with the data.

\newenvironment{acknowledgement}{\relax}{\relax}
\begin{acknowledgement}
\section*{Acknowledgements}
The ALICE collaboration would like to thank G. Giacalone, J. Noronha-Hostler, M. Luzum, and J.-Y. Ollitrault for providing the results of their hydrodynamical calculations prior to publication.
\input{fa_2018-04-13.tex}
\end{acknowledgement}

\bibliographystyle{utphys}
\bibliography{biblio}{}
\newpage
\appendix
\section{The ALICE Collaboration}
\label{app:collab}
\input{Alice_Authorlist_2018-Apr-13.tex}
\end{document}

%% file: newcommands.tex
\newcommand{\pt}{\ensuremath{p_\mathrm T\,}\xspace}
\newcommand{\at}{\ensuremath{A_\mathrm T\,}\xspace}
\newcommand{\mpt}{\ensuremath{\langle p_\mathrm T \rangle}\xspace}
\newcommand{\pp}{pp\xspace}
\newcommand{\pPb}{p--Pb\xspace}
\newcommand{\pbpb}{Pb--Pb\xspace}

\newcommand{\xexe}{Xe--Xe\xspace}

\newcommand{\gevc}{\ensuremath{{\,\mathrm{GeV}}/c}\xspace}
\newcommand{\tev}{\ensuremath{\,\mathrm{TeV}}\xspace}

\newcommand{\pA}           {\mbox{p--A}}

\newcommand{\dnchdeta}{\ensuremath{\mathrm{d}N_{\mathrm{ch}}/\mathrm{d}\eta}\xspace}

\newcommand{\sqrsn}{\ensuremath{\sqrt{s_\text{NN}}}\xspace}
\newcommand{\s}            {\ensuremath{\sqrt{s}}\xspace}
\newcommand{\sqrs}{\ensuremath{\sqrt{s}}\xspace}

\newcommand{\avg}[1]       {\ensuremath{\left\langle#1\right\rangle}}

\newcommand{\Naa}{\ensuremath{N_{\mathrm{AA}}}\xspace}

\newcommand{\raa}{\ensuremath{R_{\mathrm{AA}}}\xspace}
\newcommand{\Taa}{\ensuremath{T_{\mathrm{AA}}}\xspace}

\newcommand{\Npart}{\ensuremath{N_{\mathrm{part}}}\xspace}
\newcommand{\Ncoll}{\ensuremath{N_{\mathrm{coll}}}\xspace}
\newcommand{\avNpart}{\ensuremath{\langle N_\mathrm{part} \rangle}\xspace}
\newcommand{\avNcoll}{\ensuremath{\langle N_\mathrm{coll} \rangle}\xspace}
\newcommand{\avTAA}{\ensuremath{\langle T_\mathrm{AA} \rangle}\xspace}

\newcommand{\com}[1]       {}

\renewcommand{\xout}[1]    {}
\newcommand{\old}[1]       {\relax}

%% file: fa_2018-04-13.tex

The ALICE Collaboration would like to thank all its engineers and technicians for their invaluable contributions to the construction of the experiment and the CERN accelerator teams for the outstanding performance of the LHC complex.
The ALICE Collaboration gratefully acknowledges the resources and support provided by all Grid centres and the Worldwide LHC Computing Grid (WLCG) collaboration.
The ALICE Collaboration acknowledges the following funding agencies for their support in building and running the ALICE detector:
A. I. Alikhanyan National Science Laboratory (Yerevan Physics Institute) Foundation (ANSL), State Committee of Science and World Federation of Scientists (WFS), Armenia;
Austrian Academy of Sciences and Nationalstiftung f\"{u}r Forschung, Technologie und Entwicklung, Austria;
Ministry of Communications and High Technologies, National Nuclear Research Center, Azerbaijan;
Conselho Nacional de Desenvolvimento Cient\'{\i}fico e Tecnol\'{o}gico (CNPq), Universidade Federal do Rio Grande do Sul (UFRGS), Financiadora de Estudos e Projetos (Finep) and Funda\c{c}\~{a}o de Amparo \`{a} Pesquisa do Estado de S\~{a}o Paulo (FAPESP), Brazil;
Ministry of Science \& Technology of China (MSTC), National Natural Science Foundation of China (NSFC) and Ministry of Education of China (MOEC) , China;
Ministry of Science and Education, Croatia;
Ministry of Education, Youth and Sports of the Czech Republic, Czech Republic;
The Danish Council for Independent Research | Natural Sciences, the Carlsberg Foundation and Danish National Research Foundation (DNRF), Denmark;
Helsinki Institute of Physics (HIP), Finland;
Commissariat \`{a} l'Energie Atomique (CEA) and Institut National de Physique Nucl\'{e}aire et de Physique des Particules (IN2P3) and Centre National de la Recherche Scientifique (CNRS), France;
Bundesministerium f\"{u}r Bildung, Wissenschaft, Forschung und Technologie (BMBF) and GSI Helmholtzzentrum f\"{u}r Schwerionenforschung GmbH, Germany;
General Secretariat for Research and Technology, Ministry of Education, Research and Religions, Greece;
National Research, Development and Innovation Office, Hungary;
Department of Atomic Energy Government of India (DAE), Department of Science and Technology, Government of India (DST), University Grants Commission, Government of India (UGC) and Council of Scientific and Industrial Research (CSIR), India;
Indonesian Institute of Science, Indonesia;
Centro Fermi - Museo Storico della Fisica e Centro Studi e Ricerche Enrico Fermi and Istituto Nazionale di Fisica Nucleare (INFN), Italy;
Institute for Innovative Science and Technology , Nagasaki Institute of Applied Science (IIST), Japan Society for the Promotion of Science (JSPS) KAKENHI and Japanese Ministry of Education, Culture, Sports, Science and Technology (MEXT), Japan;
Consejo Nacional de Ciencia (CONACYT) y Tecnolog\'{i}a, through Fondo de Cooperaci\'{o}n Internacional en Ciencia y Tecnolog\'{i}a (FONCICYT) and Direcci\'{o}n General de Asuntos del Personal Academico (DGAPA), Mexico;
Nederlandse Organisatie voor Wetenschappelijk Onderzoek (NWO), Netherlands;
The Research Council of Norway, Norway;
Commission on Science and Technology for Sustainable Development in the South (COMSATS), Pakistan;
Pontificia Universidad Cat\'{o}lica del Per\'{u}, Peru;
Ministry of Science and Higher Education and National Science Centre, Poland;
Korea Institute of Science and Technology Information and National Research Foundation of Korea (NRF), Republic of Korea;
Ministry of Education and Scientific Research, Institute of Atomic Physics and Romanian National Agency for Science, Technology and Innovation, Romania;
Joint Institute for Nuclear Research (JINR), Ministry of Education and Science of the Russian Federation and National Research Centre Kurchatov Institute, Russia;
Ministry of Education, Science, Research and Sport of the Slovak Republic, Slovakia;
National Research Foundation of South Africa, South Africa;
Centro de Aplicaciones Tecnol\'{o}gicas y Desarrollo Nuclear (CEADEN), Cubaenerg\'{\i}a, Cuba and Centro de Investigaciones Energ\'{e}ticas, Medioambientales y Tecnol\'{o}gicas (CIEMAT), Spain;
Swedish Research Council (VR) and Knut \& Alice Wallenberg Foundation (KAW), Sweden;
European Organization for Nuclear Research, Switzerland;
National Science and Technology Development Agency (NSDTA), Suranaree University of Technology (SUT) and Office of the Higher Education Commission under NRU project of Thailand, Thailand;
Turkish Atomic Energy Agency (TAEK), Turkey;
National Academy of  Sciences of Ukraine, Ukraine;
Science and Technology Facilities Council (STFC), United Kingdom;
National Science Foundation of the United States of America (NSF) and United States Department of Energy, Office of Nuclear Physics (DOE NP), United States of America.

%% file: Alice_Authorlist_2018-Apr-13.tex

\begingroup
\small
\begin{flushleft}
S.~Acharya\Irefn{org138}\And 
F.T.-.~Acosta\Irefn{org22}\And 
D.~Adamov\'{a}\Irefn{org94}\And 
J.~Adolfsson\Irefn{org81}\And 
M.M.~Aggarwal\Irefn{org98}\And 
G.~Aglieri Rinella\Irefn{org36}\And 
M.~Agnello\Irefn{org33}\And 
N.~Agrawal\Irefn{org49}\And 
Z.~Ahammed\Irefn{org138}\And 
S.U.~Ahn\Irefn{org77}\And 
S.~Aiola\Irefn{org143}
% ...
\And
\renewcommand\labelenumi{\textsuperscript{\theenumi}~}

\section*{Affiliation notes}
\renewcommand\theenumi{\roman{enumi}}
\begin{Authlist}
\item \Adef{orgI}Dipartimento DET del Politecnico di Torino, Turin, Italy
\item \Adef{orgII}M.V. Lomonosov Moscow State University, D.V. Skobeltsyn Institute of Nuclear, Physics, Moscow, Russia
\item \Adef{orgIII}Department of Applied Physics, Aligarh Muslim University, Aligarh, India
\item \Adef{orgIV}Institute of Theoretical Physics, University of Wroclaw, Poland
\end{Authlist}

\section*{Collaboration Institutes}
\renewcommand\theenumi{\arabic{enumi}~}
\begin{Authlist}
\item \Idef{org1}A.I. Alikhanyan National Science Laboratory (Yerevan Physics Institute) Foundation, Yerevan, Armenia
\item \Idef{org2}Benem\'{e}rita Universidad Aut\'{o}noma de Puebla, Puebla, Mexico
\item \Idef{org3}Bogolyubov Institute for Theoretical Physics, National Academy of Sciences of Ukraine, Kiev, Ukraine
\item \Idef{org4}Bose Institute, Department of Physics  and Centre for Astroparticle Physics and Space Science (CAPSS), Kolkata, India
\item \Idef{org5}Budker Institute for Nuclear Physics, Novosibirsk, Russia
\item \Idef{org6}California Polytechnic State University, San Luis Obispo, California, United States
\item \Idef{org7}Central China Normal University, Wuhan, China
\item \Idef{org8}Centre de Calcul de l'IN2P3, Villeurbanne, Lyon, France
\item \Idef{org9}Centro de Aplicaciones Tecnol\'{o}gicas y Desarrollo Nuclear (CEADEN), Havana, Cuba
\item \Idef{org10}Centro de Investigaci\'{o}n y de Estudios Avanzados (CINVESTAV), Mexico City and M\'{e}rida, Mexico
\item \Idef{org11}Centro Fermi - Museo Storico della Fisica e Centro Studi e Ricerche ``Enrico Fermi', Rome, Italy
\item \Idef{org12}Chicago State University, Chicago, Illinois, United States
\item \Idef{org13}China Institute of Atomic Energy, Beijing, China
\item \Idef{org14}Chonbuk National University, Jeonju, Republic of Korea
\item \Idef{org15}Comenius University Bratislava, Faculty of Mathematics, Physics and Informatics, Bratislava, Slovakia
\item \Idef{org16}COMSATS Institute of Information Technology (CIIT), Islamabad, Pakistan
\item \Idef{org17}Creighton University, Omaha, Nebraska, United States
\item \Idef{org18}Department of Physics, Aligarh Muslim University, Aligarh, India
\item \Idef{org19}Department of Physics, Ohio State University, Columbus, Ohio, United States
\item \Idef{org20}Department of Physics, Pusan National University, Pusan, Republic of Korea
\item \Idef{org21}Department of Physics, Sejong University, Seoul, Republic of Korea
\item \Idef{org22}Department of Physics, University of California, Berkeley, California, United States
\item \Idef{org23}Department of Physics, University of Oslo, Oslo, Norway
\item \Idef{org24}Department of Physics and Technology, University of Bergen, Bergen, Norway
\item \Idef{org25}Dipartimento di Fisica dell'Universit\`{a} 'La Sapienza' and Sezione INFN, Rome, Italy
\item \Idef{org26}Dipartimento di Fisica dell'Universit\`{a} and Sezione INFN, Cagliari, Italy
\item \Idef{org27}Dipartimento di Fisica dell'Universit\`{a} and Sezione INFN, Trieste, Italy
\item \Idef{org28}Dipartimento di Fisica dell'Universit\`{a} and Sezione INFN, Turin, Italy
\item \Idef{org29}Dipartimento di Fisica e Astronomia dell'Universit\`{a} and Sezione INFN, Bologna, Italy
\item \Idef{org30}Dipartimento di Fisica e Astronomia dell'Universit\`{a} and Sezione INFN, Catania, Italy
\item \Idef{org31}Dipartimento di Fisica e Astronomia dell'Universit\`{a} and Sezione INFN, Padova, Italy
\item \Idef{org32}Dipartimento di Fisica `E.R.~Caianiello' dell'Universit\`{a} and Gruppo Collegato INFN, Salerno, Italy
\item \Idef{org33}Dipartimento DISAT del Politecnico and Sezione INFN, Turin, Italy
\item \Idef{org34}Dipartimento di Scienze e Innovazione Tecnologica dell'Universit\`{a} del Piemonte Orientale and INFN Sezione di Torino, Alessandria, Italy
\item \Idef{org35}Dipartimento Interateneo di Fisica `M.~Merlin' and Sezione INFN, Bari, Italy
\item \Idef{org36}European Organization for Nuclear Research (CERN), Geneva, Switzerland
\item \Idef{org37}Faculty of Electrical Engineering, Mechanical Engineering and Naval Architecture, University of Split, Split, Croatia
\item \Idef{org38}Faculty of Engineering and Science, Western Norway University of Applied Sciences, Bergen, Norway
\item \Idef{org39}Faculty of Nuclear Sciences and Physical Engineering, Czech Technical University in Prague, Prague, Czech Republic
\item \Idef{org40}Faculty of Science, P.J.~\v{S}af\'{a}rik University, Ko\v{s}ice, Slovakia
\item \Idef{org41}Frankfurt Institute for Advanced Studies, Johann Wolfgang Goethe-Universit\"{a}t Frankfurt, Frankfurt, Germany
\item \Idef{org42}Gangneung-Wonju National University, Gangneung, Republic of Korea
\item \Idef{org43}Gauhati University, Department of Physics, Guwahati, India
\item \Idef{org44}Helmholtz-Institut f\"{u}r Strahlen- und Kernphysik, Rheinische Friedrich-Wilhelms-Universit\"{a}t Bonn, Bonn, Germany
\item \Idef{org45}Helsinki Institute of Physics (HIP), Helsinki, Finland
\item \Idef{org46}Hiroshima University, Hiroshima, Japan
\item \Idef{org47}Hochschule Worms, Zentrum  f\"{u}r Technologietransfer und Telekommunikation (ZTT), Worms, Germany
\item \Idef{org48}Horia Hulubei National Institute of Physics and Nuclear Engineering, Bucharest, Romania
\item \Idef{org49}Indian Institute of Technology Bombay (IIT), Mumbai, India
\item \Idef{org50}Indian Institute of Technology Indore, Indore, India
\item \Idef{org51}Indonesian Institute of Sciences, Jakarta, Indonesia
\item \Idef{org52}INFN, Laboratori Nazionali di Frascati, Frascati, Italy
\item \Idef{org53}INFN, Sezione di Bari, Bari, Italy
\item \Idef{org54}INFN, Sezione di Bologna, Bologna, Italy
\item \Idef{org55}INFN, Sezione di Cagliari, Cagliari, Italy
\item \Idef{org56}INFN, Sezione di Catania, Catania, Italy
\item \Idef{org57}INFN, Sezione di Padova, Padova, Italy
\item \Idef{org58}INFN, Sezione di Roma, Rome, Italy
\item \Idef{org59}INFN, Sezione di Torino, Turin, Italy
\item \Idef{org60}INFN, Sezione di Trieste, Trieste, Italy
\item \Idef{org61}Inha University, Incheon, Republic of Korea
\item \Idef{org62}Institut de Physique Nucl\'{e}aire d'Orsay (IPNO), Institut National de Physique Nucl\'{e}aire et de Physique des Particules (IN2P3/CNRS), Universit\'{e} de Paris-Sud, Universit\'{e} Paris-Saclay, Orsay, France
\item \Idef{org63}Institute for Nuclear Research, Academy of Sciences, Moscow, Russia
\item \Idef{org64}Institute for Subatomic Physics, Utrecht University/Nikhef, Utrecht, Netherlands
\item \Idef{org65}Institute for Theoretical and Experimental Physics, Moscow, Russia
\item \Idef{org66}Institute of Experimental Physics, Slovak Academy of Sciences, Ko\v{s}ice, Slovakia
\item \Idef{org67}Institute of Physics, Bhubaneswar, India
\item \Idef{org68}Institute of Physics of the Czech Academy of Sciences, Prague, Czech Republic
\item \Idef{org69}Institute of Space Science (ISS), Bucharest, Romania
\item \Idef{org70}Institut f\"{u}r Kernphysik, Johann Wolfgang Goethe-Universit\"{a}t Frankfurt, Frankfurt, Germany
\item \Idef{org71}Instituto de Ciencias Nucleares, Universidad Nacional Aut\'{o}noma de M\'{e}xico, Mexico City, Mexico
\item \Idef{org72}Instituto de F\'{i}sica, Universidade Federal do Rio Grande do Sul (UFRGS), Porto Alegre, Brazil
\item \Idef{org73}Instituto de F\'{\i}sica, Universidad Nacional Aut\'{o}noma de M\'{e}xico, Mexico City, Mexico
\item \Idef{org74}iThemba LABS, National Research Foundation, Somerset West, South Africa
\item \Idef{org75}Johann-Wolfgang-Goethe Universit\"{a}t Frankfurt Institut f\"{u}r Informatik, Fachbereich Informatik und Mathematik, Frankfurt, Germany
\item \Idef{org76}Joint Institute for Nuclear Research (JINR), Dubna, Russia
\item \Idef{org77}Korea Institute of Science and Technology Information, Daejeon, Republic of Korea
\item \Idef{org78}KTO Karatay University, Konya, Turkey
\item \Idef{org79}Laboratoire de Physique Subatomique et de Cosmologie, Universit\'{e} Grenoble-Alpes, CNRS-IN2P3, Grenoble, France
\item \Idef{org80}Lawrence Berkeley National Laboratory, Berkeley, California, United States
\item \Idef{org81}Lund University Department of Physics, Division of Particle Physics, Lund, Sweden
\item \Idef{org82}Nagasaki Institute of Applied Science, Nagasaki, Japan
\item \Idef{org83}Nara Women{'}s University (NWU), Nara, Japan
\item \Idef{org84}National and Kapodistrian University of Athens, School of Science, Department of Physics , Athens, Greece
\item \Idef{org85}National Centre for Nuclear Research, Warsaw, Poland
\item \Idef{org86}National Institute of Science Education and Research, HBNI, Jatni, India
\item \Idef{org87}National Nuclear Research Center, Baku, Azerbaijan
\item \Idef{org88}National Research Centre Kurchatov Institute, Moscow, Russia
\item \Idef{org89}Niels Bohr Institute, University of Copenhagen, Copenhagen, Denmark
\item \Idef{org90}Nikhef, National institute for subatomic physics, Amsterdam, Netherlands
\item \Idef{org91}NRC Kurchatov Institute IHEP, Protvino, Russia
\item \Idef{org92}NRNU Moscow Engineering Physics Institute, Moscow, Russia
\item \Idef{org93}Nuclear Physics Group, STFC Daresbury Laboratory, Daresbury, United Kingdom
\item \Idef{org94}Nuclear Physics Institute of the Czech Academy of Sciences, \v{R}e\v{z} u Prahy, Czech Republic
\item \Idef{org95}Oak Ridge National Laboratory, Oak Ridge, Tennessee, United States
\item \Idef{org96}Petersburg Nuclear Physics Institute, Gatchina, Russia
\item \Idef{org97}Physics department, Faculty of science, University of Zagreb, Zagreb, Croatia
\item \Idef{org98}Physics Department, Panjab University, Chandigarh, India
\item \Idef{org99}Physics Department, University of Jammu, Jammu, India
\item \Idef{org100}Physics Department, University of Rajasthan, Jaipur, India
\item \Idef{org101}Physikalisches Institut, Eberhard-Karls-Universit\"{a}t T\"{u}bingen, T\"{u}bingen, Germany
\item \Idef{org102}Physikalisches Institut, Ruprecht-Karls-Universit\"{a}t Heidelberg, Heidelberg, Germany
\item \Idef{org103}Physik Department, Technische Universit\"{a}t M\"{u}nchen, Munich, Germany
\item \Idef{org104}Research Division and ExtreMe Matter Institute EMMI, GSI Helmholtzzentrum f\"ur Schwerionenforschung GmbH, Darmstadt, Germany
\item \Idef{org105}Rudjer Bo\v{s}kovi\'{c} Institute, Zagreb, Croatia
\item \Idef{org106}Russian Federal Nuclear Center (VNIIEF), Sarov, Russia
\item \Idef{org107}Saha Institute of Nuclear Physics, Kolkata, India
\item \Idef{org108}School of Physics and Astronomy, University of Birmingham, Birmingham, United Kingdom
\item \Idef{org109}Secci\'{o}n F\'{\i}sica, Departamento de Ciencias, Pontificia Universidad Cat\'{o}lica del Per\'{u}, Lima, Peru
\item \Idef{org110}Shanghai Institute of Applied Physics, Shanghai, China
\item \Idef{org111}Stefan Meyer Institut f\"{u}r Subatomare Physik (SMI), Vienna, Austria
\item \Idef{org112}SUBATECH, IMT Atlantique, Universit\'{e} de Nantes, CNRS-IN2P3, Nantes, France
\item \Idef{org113}Suranaree University of Technology, Nakhon Ratchasima, Thailand
\item \Idef{org114}Technical University of Ko\v{s}ice, Ko\v{s}ice, Slovakia
\item \Idef{org115}Technische Universit\"{a}t M\"{u}nchen, Excellence Cluster 'Universe', Munich, Germany
\item \Idef{org116}The Henryk Niewodniczanski Institute of Nuclear Physics, Polish Academy of Sciences, Cracow, Poland
\item \Idef{org117}The University of Texas at Austin, Austin, Texas, United States
\item \Idef{org118}Universidad Aut\'{o}noma de Sinaloa, Culiac\'{a}n, Mexico
\item \Idef{org119}Universidade de S\~{a}o Paulo (USP), S\~{a}o Paulo, Brazil
\item \Idef{org120}Universidade Estadual de Campinas (UNICAMP), Campinas, Brazil
\item \Idef{org121}Universidade Federal do ABC, Santo Andre, Brazil
\item \Idef{org122}University College of Southeast Norway, Tonsberg, Norway
\item \Idef{org123}University of Cape Town, Cape Town, South Africa
\item \Idef{org124}University of Houston, Houston, Texas, United States
\item \Idef{org125}University of Jyv\"{a}skyl\"{a}, Jyv\"{a}skyl\"{a}, Finland
\item \Idef{org126}University of Liverpool, Department of Physics Oliver Lodge Laboratory , Liverpool, United Kingdom
\item \Idef{org127}University of Tennessee, Knoxville, Tennessee, United States
\item \Idef{org128}University of the Witwatersrand, Johannesburg, South Africa
\item \Idef{org129}University of Tokyo, Tokyo, Japan
\item \Idef{org130}University of Tsukuba, Tsukuba, Japan
\item \Idef{org131}Universit\'{e} Clermont Auvergne, CNRS/IN2P3, LPC, Clermont-Ferrand, France
\item \Idef{org132}Universit\'{e} de Lyon, Universit\'{e} Lyon 1, CNRS/IN2P3, IPN-Lyon, Villeurbanne, Lyon, France
\item \Idef{org133}Universit\'{e} de Strasbourg, CNRS, IPHC UMR 7178, F-67000 Strasbourg, France, Strasbourg, France
\item \Idef{org134} Universit\'{e} Paris-Saclay Centre d¿\'Etudes de Saclay (CEA), IRFU, Department de Physique Nucl\'{e}aire (DPhN), Saclay, France
\item \Idef{org135}Universit\`{a} degli Studi di Pavia, Pavia, Italy
\item \Idef{org136}Universit\`{a} di Brescia, Brescia, Italy
\item \Idef{org137}V.~Fock Institute for Physics, St. Petersburg State University, St. Petersburg, Russia
\item \Idef{org138}Variable Energy Cyclotron Centre, Kolkata, India
\item \Idef{org139}Warsaw University of Technology, Warsaw, Poland
\item \Idef{org140}Wayne State University, Detroit, Michigan, United States
\item \Idef{org141}Westf\"{a}lische Wilhelms-Universit\"{a}t M\"{u}nster, Institut f\"{u}r Kernphysik, M\"{u}nster, Germany
\item \Idef{org142}Wigner Research Centre for Physics, Hungarian Academy of Sciences, Budapest, Hungary
\item \Idef{org143}Yale University, New Haven, Connecticut, United States
\item \Idef{org144}Yonsei University, Seoul, Republic of Korea
\end{Authlist}
\endgroup

%% file: pmain.bbl
\providecommand{\href}[2]{#2}\begingroup\raggedright\begin{thebibliography}{10}

\bibitem{Muller:2012zq}
B.~Muller, J.~Schukraft, and B.~Wyslouch, ``{First Results from Pb+Pb
  collisions at the LHC},''
  \href{http://dx.doi.org/10.1146/annurev-nucl-102711-094910}{{\em Ann. Rev.
  Nucl. Part. Sci.} {\bfseries 62} (2012) 361--386},
\href{http://arxiv.org/abs/1202.3233}{{\ttfamily arXiv:1202.3233 [hep-ex]}}.

\bibitem{Heinz:2013th}
U.~Heinz and R.~Snellings, ``{Collective flow and viscosity in relativistic
  heavy-ion collisions},''
  \href{http://dx.doi.org/10.1146/annurev-nucl-102212-170540}{{\em Ann. Rev.
  Nucl. Part. Sci.} {\bfseries 63} (2013) 123--151},
\href{http://arxiv.org/abs/1301.2826}{{\ttfamily arXiv:1301.2826 [nucl-th]}}.

\bibitem{Bernhard:2016tnd}
J.~E. Bernhard, J.~S. Moreland, S.~A. Bass, J.~Liu, and U.~Heinz, ``{Applying
  Bayesian parameter estimation to relativistic heavy-ion collisions:
  simultaneous characterization of the initial state and quark-gluon plasma
  medium},'' \href{http://dx.doi.org/10.1103/PhysRevC.94.024907}{{\em Phys.
  Rev.} {\bfseries C94} no.~2, (2016) 024907},
\href{http://arxiv.org/abs/1605.03954}{{\ttfamily arXiv:1605.03954 [nucl-th]}}.

\bibitem{Adam:2016tre}
{\bfseries ALICE} Collaboration, J.~Adam {\em et~al.}, ``{Centrality dependence
  of the nuclear modification factor of charged pions, kaons, and protons in
  Pb--Pb collisions at $\snn = 2.76$~TeV},''
  \href{http://dx.doi.org/10.1103/PhysRevC.93.034913}{{\em Phys. Rev. C}
  {\bfseries 93} (2016) 034913},
\href{http://arxiv.org/abs/1506.07287}{{\ttfamily arXiv:1506.07287 [nucl-ex]}}.

\bibitem{Acharya:2018qsh}
{\bfseries ALICE} Collaboration, S.~Acharya {\em et~al.}, ``{Transverse
  momentum spectra and nuclear modification factors of charged particles in pp,
  p-Pb and Pb-Pb collisions at the LHC},''
\href{http://arxiv.org/abs/1802.09145}{{\ttfamily arXiv:1802.09145 [nucl-ex]}}.

\bibitem{Adam:2016dau}
{\bfseries ALICE} Collaboration, J.~Adam {\em et~al.}, ``{Multiplicity
  dependence of charged pion, kaon, and (anti)proton production at large
  transverse momentum in p--Pb Collisions at $\sqrt{s_{_{\rm NN}}}$ = 5.02
  TeV},'' \href{http://dx.doi.org/10.1016/j.physletb.2016.07.050}{{\em Phys.
  Lett. B} {\bfseries 760} (2016) 720--735},
\href{http://arxiv.org/abs/1601.03658}{{\ttfamily arXiv:1601.03658 [nucl-ex]}}.

\bibitem{Ortiz:2013yxa}
A.~Ortiz, P.~Christiansen, E.~Cuautle, I.~Maldonado, and G.~Paic, ``{Color
  reconnection and flow-like patterns in pp collisions},''
  \href{http://dx.doi.org/10.1103/PhysRevLett.111.042001}{{\em Phys. Rev.
  Lett.} {\bfseries 111} no.~4, (2013) 042001},
\href{http://arxiv.org/abs/1303.6326}{{\ttfamily arXiv:1303.6326 [hep-ph]}}.

\bibitem{ABELEV:2013wsa}
{\bfseries ALICE} Collaboration, B.~B. Abelev {\em et~al.}, ``{Long-range
  angular correlations of pi, K and p in p--Pb collisions at $\sqrt{s_{_{\rm
  NN}}}=5.02$ TeV},''
  \href{http://dx.doi.org/10.1016/j.physletb.2013.08.024}{{\em Phys. Lett. B}
  {\bfseries 726} (2013) 164--177},
\href{http://arxiv.org/abs/1307.3237}{{\ttfamily arXiv:1307.3237 [nucl-ex]}}.

\bibitem{Khachatryan:2014jra}
{\bfseries CMS} Collaboration, V.~Khachatryan {\em et~al.}, ``{Long-range
  two-particle correlations of strange hadrons with charged particles in pPb
  and PbPb collisions at LHC energies},''
  \href{http://dx.doi.org/10.1016/j.physletb.2015.01.034}{{\em Phys. Lett.}
  {\bfseries B742} (2015) 200--224},
\href{http://arxiv.org/abs/1409.3392}{{\ttfamily arXiv:1409.3392 [nucl-ex]}}.

\bibitem{Chatrchyan:2013nka}
{\bfseries CMS} Collaboration, S.~Chatrchyan {\em et~al.}, ``{Multiplicity and
  transverse-momentum dependence of two- and four-particle correlations in pPb
  and PbPb collisions},''
  \href{http://dx.doi.org/10.1016/j.physletb.2013.06.028}{{\em Phys. Lett. B}
  {\bfseries 724} (2013) 213--240},
\href{http://arxiv.org/abs/1305.0609}{{\ttfamily arXiv:1305.0609 [nucl-ex]}}.

\bibitem{Khachatryan:2015waa}
{\bfseries CMS} Collaboration, V.~Khachatryan {\em et~al.}, ``{Evidence for
  Collective Multiparticle Correlations in p-Pb Collisions},''
  \href{http://dx.doi.org/10.1103/PhysRevLett.115.012301}{{\em Phys. Rev.
  Lett.} {\bfseries 115} no.~1, (2015) 012301},
\href{http://arxiv.org/abs/1502.05382}{{\ttfamily arXiv:1502.05382 [nucl-ex]}}.

\bibitem{Aad:2013fja}
{\bfseries ATLAS} Collaboration, G.~Aad {\em et~al.}, ``{Measurement with the
  ATLAS detector of multi-particle azimuthal correlations in p+Pb collisions at
  $\sqrt{s_{_{\rm NN}}}=5.02$ TeV},''
  \href{http://dx.doi.org/10.1016/j.physletb.2013.06.057}{{\em Phys. Lett. B}
  {\bfseries 725} (2013) 60--78},
\href{http://arxiv.org/abs/1303.2084}{{\ttfamily arXiv:1303.2084 [hep-ex]}}.

\bibitem{Dusling:2013oia}
K.~Dusling and R.~Venugopalan, ``{Comparison of the Color Glass Condensate to
  di-hadron correlations in proton-proton and proton-nucleus collisions},''
  \href{http://dx.doi.org/10.1103/PhysRevD.87.094034}{{\em Phys. Rev. D}
  {\bfseries 87} (2013) 094034},
\href{http://arxiv.org/abs/1302.7018}{{\ttfamily arXiv:1302.7018 [hep-ph]}}.

\bibitem{Blok:2017pui}
B.~Blok, C.~D. Jakel, M.~Strikman, and U.~A. Wiedemann, ``{Collectivity from
  interference},'' \href{http://dx.doi.org/10.1007/JHEP12(2017)074}{{\em JHEP}
  {\bfseries 12} (2017) 074},
\href{http://arxiv.org/abs/1708.08241}{{\ttfamily arXiv:1708.08241 [hep-ph]}}.

\bibitem{Adare:2015bua}
{\bfseries PHENIX} Collaboration, A.~Adare {\em et~al.}, ``{Transverse energy
  production and charged-particle multiplicity at midrapidity in various
  systems from $\sqrt{s_{NN}}=7.7$ to 200 GeV},''
  \href{http://dx.doi.org/10.1103/PhysRevC.93.024901}{{\em Phys. Rev.}
  {\bfseries C93} no.~2, (2016) 024901},
\href{http://arxiv.org/abs/1509.06727}{{\ttfamily arXiv:1509.06727 [nucl-ex]}}.

\bibitem{Adam:2016thv}
{\bfseries ALICE} Collaboration, J.~Adam {\em et~al.}, ``{Measurement of
  transverse energy at midrapidity in Pb-Pb collisions at $\sqrt{s_{\rm NN}} =
  2.76$ TeV},'' \href{http://dx.doi.org/10.1103/PhysRevC.94.034903}{{\em Phys.
  Rev.} {\bfseries C94} no.~3, (2016) 034903},
\href{http://arxiv.org/abs/1603.04775}{{\ttfamily arXiv:1603.04775 [nucl-ex]}}.

\bibitem{Adam:2015vna}
{\bfseries ALICE} Collaboration, J.~Adam {\em et~al.}, ``{Centrality dependence
  of pion freeze-out radii in Pb-Pb collisions at $\sqrt{s_{\rm NN}} = 2.76$
  TeV},'' \href{http://dx.doi.org/10.1103/PhysRevC.93.024905}{{\em Phys. Rev.}
  {\bfseries C93} no.~2, (2016) 024905},
\href{http://arxiv.org/abs/1507.06842}{{\ttfamily arXiv:1507.06842 [nucl-ex]}}.

\bibitem{ALICEpubCent}
{\bfseries ALICE Collaboration} Collaboration, ``{Centrality determination
  using the Glauber model in Xe-Xe collisions at $\sqrt{s_{\rm NN}} = 5.44$
  TeV},''. \url{https://cds.cern.ch/record/2315401}.

\bibitem{xexe-cent}
{\bfseries ALICE} Collaboration, S.~Acharya {\em et~al.}, ``{Centrality and
  pseudorapidity dependence of the charged-particle multiplicity density in
  Xe-Xe collisions at $\sqrt{s_{\rm NN}}$ = 5.44 TeV},''
\href{http://arxiv.org/abs/1805.04432}{{\ttfamily arXiv:1805.04432 [nucl-ex]}}.

\bibitem{Alver:2005nb}
{\bfseries PHOBOS} Collaboration, B.~Alver {\em et~al.}, ``{System size and
  centrality dependence of charged hadron transverse momentum spectra in Au +
  Au and Cu + Cu collisions at s(NN)**(1/2) = 62.4-GeV and 200-GeV},''
  \href{http://dx.doi.org/10.1103/PhysRevLett.96.212301}{{\em Phys. Rev. Lett.}
  {\bfseries 96} (2006) 212301},
\href{http://arxiv.org/abs/nucl-ex/0512016}{{\ttfamily arXiv:nucl-ex/0512016
  [nucl-ex]}}.

\bibitem{Bjorken:1982tu}
J.~Bjorken, ``{Energy loss of energetic Partons in quark - gluon plasma:
  possible extinction of high $p_{\mathrm T}$ jets in hadron - hadron
  collisions},'' Tech. Rep. {FERMILAB-PUB-82-059-T}, Fermilab, 1982.
\newblock
  \url{http://lss.fnal.gov/archive/preprint/fermilab-pub-82-059-t.shtml}.

\bibitem{d'Enterria:2009am}
D.~d'Enterria, ``{Jet quenching},'' {\em Landolt-B{\" o}rnstein} {\bfseries
  I/23} (2010) ,
\href{http://arxiv.org/abs/0902.2011}{{\ttfamily arXiv:0902.2011 [nucl-ex]}}.

\bibitem{Ortiz:2017cul}
A.~Ortiz and O.~V\'azquez, ``{Energy density and path-length dependence of the
  fractional momentum loss in heavy-ion collisions at $\sqrt{s_{\rm NN}}$ from
  62.4 to 5020 GeV},'' \href{http://dx.doi.org/10.1103/PhysRevC.97.014910}{{\em
  Phys. Rev.} {\bfseries C97} no.~1, (2018) 014910},
\href{http://arxiv.org/abs/1708.07571}{{\ttfamily arXiv:1708.07571 [hep-ph]}}.

\bibitem{Loizides:2017ack}
C.~Loizides, J.~Kamin, and D.~d'Enterria, ``{Precision Monte Carlo Glauber
  predictions at present and future nuclear colliders},''
\href{http://arxiv.org/abs/1710.07098}{{\ttfamily arXiv:1710.07098 [nucl-ex]}}.

\bibitem{Khachatryan:2016odn}
{\bfseries CMS} Collaboration, V.~Khachatryan {\em et~al.}, ``{Charged-particle
  nuclear modification factors in PbPb and pPb collisions at
  $\sqrt{s_{\mathrm{NN}}}= $ 5.02 TeV},''
  \href{http://dx.doi.org/10.1007/JHEP04(2017)039}{{\em JHEP} {\bfseries 04}
  (2017) 039},
\href{http://arxiv.org/abs/1611.01664}{{\ttfamily arXiv:1611.01664 [nucl-ex]}}.

\bibitem{DEVRIES1987495}
H.~D. Vries, C.~D. Jager, and C.~D. Vries, ``Nuclear
  charge-density-distribution parameters from elastic electron scattering,''
  \href{http://dx.doi.org/http://dx.doi.org/10.1016/0092-640X(87)90013-1}{{\em
  Atomic Data and Nuclear Data Tables} {\bfseries 36} (1987) 495--536}.
  \url{http://www.sciencedirect.com/science/article/pii/0092640X87900131}.

\bibitem{Aamodt:2008zz}
{\bfseries ALICE} Collaboration, K.~Aamodt {\em et~al.}, ``{The ALICE
  experiment at the CERN LHC},''
\href{http://dx.doi.org/10.1088/1748-0221/3/08/S08002}{{\em JINST} {\bfseries
  3} (2008) S08002}.

\bibitem{Aamodt:2010aa}
{\bfseries ALICE} Collaboration, K.~Aamodt {\em et~al.}, ``{Alignment of the
  ALICE Inner Tracking System with cosmic-ray tracks},''
  \href{http://dx.doi.org/10.1088/1748-0221/5/03/P03003}{{\em JINST} {\bfseries
  5} (2010) P03003},
\href{http://arxiv.org/abs/1001.0502}{{\ttfamily arXiv:1001.0502
  [physics.ins-det]}}.

\bibitem{Alme:2010ke}
J.~Alme {\em et~al.}, ``{The ALICE TPC, a large 3-dimensional tracking device
  with fast readout for ultra-high multiplicity events},''
  \href{http://dx.doi.org/10.1016/j.nima.2010.04.042}{{\em Nucl. Instrum. Meth.
  A} {\bfseries 622} (2010) 316--367},
\href{http://arxiv.org/abs/1001.1950}{{\ttfamily arXiv:1001.1950
  [physics.ins-det]}}.

\bibitem{Adam:2015ptt}
{\bfseries ALICE} Collaboration, J.~Adam {\em et~al.}, ``{Centrality dependence
  of the charged-particle multiplicity density at midrapidity in Pb--Pb
  collisions at $\sqrt{s_{\rm NN}}$ = 5.02 TeV},''
  \href{http://dx.doi.org/10.1103/PhysRevLett.116.222302}{{\em Phys. Rev.
  Lett.} {\bfseries 116} (2016) 222302},
\href{http://arxiv.org/abs/1512.06104}{{\ttfamily arXiv:1512.06104 [nucl-ex]}}.

\bibitem{Acharya:2017dpp}
{\bfseries ALICE} Collaboration, S.~Acharya {\em et~al.}, ``{The ALICE
  definition of primary particles},'' Public Note {ALICE-PUBLIC-2017-005},
  CERN, 2017.
\newblock \url{https://cds.cern.ch/record/2270008}.

\bibitem{Wang:1991hta}
X.-N. Wang and M.~Gyulassy, ``{HIJING: A Monte Carlo model for multiple jet
  production in p--p, p--A and A--A collisions},''
\href{http://dx.doi.org/10.1103/PhysRevD.44.3501}{{\em Phys. Rev. D} {\bfseries
  44} (1991) 3501--3516}.

\bibitem{GEANT3}
R.~Brun, F.~Bruyant, F.~Carminati, S.~Giani, M.~Maire, A.~McPherson,
  G.~Patrick, and L.~Urban, {\em {GEANT: Detector Description and Simulation
  Tool; Oct 1994}}.
\newblock CERN Program Library. CERN, Geneva, 1993.
\newblock \url{https://cds.cern.ch/record/1082634}.
\newblock Long Writeup W5013.

\bibitem{ALICE:2017jyt}
{\bfseries ALICE} Collaboration, J.~Adam {\em et~al.}, ``{Enhanced production
  of multi-strange hadrons in high-multiplicity proton-proton collisions},''
  \href{http://dx.doi.org/10.1038/nphys4111}{{\em Nature Phys.} {\bfseries 13}
  (2017) 535--539},
\href{http://arxiv.org/abs/1606.07424}{{\ttfamily arXiv:1606.07424 [nucl-ex]}}.

\bibitem{Abelev:2013vea}
{\bfseries ALICE} Collaboration, B.~Abelev {\em et~al.}, ``{Centrality
  dependence of $\pi$, K, p production in Pb-Pb collisions at $\sqrt{s_{_{\rm
  NN}}}=2.76$ TeV},'' \href{http://dx.doi.org/10.1103/PhysRevC.88.044910}{{\em
  Phys. Rev. C} {\bfseries 88} (2013) 044910},
\href{http://arxiv.org/abs/1303.0737}{{\ttfamily arXiv:1303.0737 [hep-ex]}}.

\bibitem{Abelev:2013xaa}
{\bfseries ALICE} Collaboration, B.~Abelev {\em et~al.}, ``{$K^0_S$ and
  $\Lambda$ production in Pb--Pb collisions at $\sqrt{s_{_{\rm NN}}}=2.76$
  TeV},'' \href{http://dx.doi.org/10.1103/PhysRevLett.111.222301}{{\em Phys.
  Rev. Lett.} {\bfseries 111} (2013) 222301},
\href{http://arxiv.org/abs/1307.5530}{{\ttfamily arXiv:1307.5530 [nucl-ex]}}.

\bibitem{Abelev:2014laa}
{\bfseries ALICE} Collaboration, B.~Abelev {\em et~al.}, ``{Production of
  charged pions, kaons and protons at large transverse momenta in pp and Pb--Pb
  collisions at $\sqrt{s_{_{\rm NN}}}=2.76$ TeV},''
  \href{http://dx.doi.org/10.1016/j.physletb.2014.07.011}{{\em Phys. Lett. B}
  {\bfseries 736} (2014) 196--207},
\href{http://arxiv.org/abs/1401.1250}{{\ttfamily arXiv:1401.1250 [nucl-ex]}}.

\bibitem{Abelev:2014ffa}
{\bfseries ALICE} Collaboration, B.~B. Abelev {\em et~al.}, ``{Performance of
  the ALICE experiment at the CERN LHC},''
  \href{http://dx.doi.org/10.1142/S0217751X14300440}{{\em Int. J. Mod. Phys. A}
  {\bfseries 29} (2014) 1430044},
\href{http://arxiv.org/abs/1402.4476}{{\ttfamily arXiv:1402.4476 [nucl-ex]}}.

\bibitem{Abelev:2013ala}
{\bfseries ALICE} Collaboration, B.~Abelev {\em et~al.}, ``{Energy dependence
  of the transverse momentum distributions of charged particles in pp
  collisions measured by ALICE},''
  \href{http://dx.doi.org/10.1140/epjc/s10052-013-2662-9}{{\em Eur. Phys. J. C}
  {\bfseries 73} (2013) 2662},
\href{http://arxiv.org/abs/1307.1093}{{\ttfamily arXiv:1307.1093 [nucl-ex]}}.

\bibitem{Skands:2014mon}
{P. Skands, S. Carrazza, J. Rojo}, ``{Tuning PYTHIA 8.1: the Monash 2013
  tune},'' \href{http://dx.doi.org/10.1140/epjc/s10052-014-3024-y}{{\em Eur.
  Phys. J. C} {\bfseries 74} (2014) 3024},
\href{http://arxiv.org/abs/1404.5630}{{\ttfamily arXiv:1404.5630 [hep-ph]}}.

\bibitem{Adare:2015cua}
{\bfseries PHENIX} Collaboration, A.~Adare {\em et~al.}, ``{Scaling properties
  of fractional momentum loss of high-$p_T$ hadrons in nucleus-nucleus
  collisions at $\sqrt{s_{_{NN}}}$ from 62.4 GeV to 2.76 TeV},''
  \href{http://dx.doi.org/10.1103/PhysRevC.93.024911}{{\em Phys. Rev.}
  {\bfseries C93} no.~2, (2016) 024911},
\href{http://arxiv.org/abs/1509.06735}{{\ttfamily arXiv:1509.06735 [nucl-ex]}}.

\bibitem{Giacalone:2017dud}
G.~Giacalone, J.~Noronha-Hostler, M.~Luzum, and J.-Y. Ollitrault,
  ``{Hydrodynamic predictions for 5.44 TeV Xe+Xe collisions},''
  \href{http://dx.doi.org/10.1103/PhysRevC.97.034904}{{\em Phys. Rev.}
  {\bfseries C97} no.~3, (2018) 034904},
\href{http://arxiv.org/abs/1711.08499}{{\ttfamily arXiv:1711.08499 [nucl-th]}}.

\bibitem{Kolb:2000fha}
P.~F. Kolb, P.~Huovinen, U.~W. Heinz, and H.~Heiselberg, ``{Elliptic flow at
  SPS and RHIC: From kinetic transport to hydrodynamics},''
  \href{http://dx.doi.org/10.1016/S0370-2693(01)00079-X}{{\em Phys. Lett.}
  {\bfseries B500} (2001) 232--240},
\href{http://arxiv.org/abs/hep-ph/0012137}{{\ttfamily arXiv:hep-ph/0012137
  [hep-ph]}}.

\bibitem{ex:PbPb:CMSTransversEnergy:2012}
{\bfseries CMS Collaboration} Collaboration, T.~C. Collaboration, ``Measurement
  of the pseudorapidity and centrality dependence of the transverse energy
  density in pb-pb collisions at $\sqrt{{s}_{\mathrm{NN}}}=2.76\text{ }\text{
  }\mathrm{TeV}$,''
  \href{http://dx.doi.org/10.1103/PhysRevLett.109.152303}{{\em Phys. Rev.
  Lett.} {\bfseries 109} (Oct, 2012) 152303}.
  \url{https://link.aps.org/doi/10.1103/PhysRevLett.109.152303}.

\bibitem{Djordjevic:2018}
M.~Djordjevic, D.~Zigic, M.~Djordjevic, and J.~Auvinen, ``{How to test
  path-length dependence in energy loss mechanisms: analysis leading to a new
  observable},''  (2018) ,
\href{http://arxiv.org/abs/1805.04030}{{\ttfamily arXiv:1805.04030 [nucl-th]}}.

\bibitem{Hagedorn:1983tyu}
R.~Hagedorn, ``{Multiplicities, $\pt$ distributions and the expected hadron
  $\to$ quark-gluon phase transition},''
\href{http://dx.doi.org/10.1007/BF02740917}{{\em Riv. Nuovo Cim.} {\bfseries 6}
  (1983) 1--50}.

\end{thebibliography}\endgroup
